\documentclass[fleqn,usenatbib]{mnras}

\usepackage{newtxtext,newtxmath}

\usepackage[T1]{fontenc}

\DeclareRobustCommand{\VAN}[3]{#2}
\let\VANthebibliography\thebibliography
\def\thebibliography{\DeclareRobustCommand{\VAN}[3]{##3}\VANthebibliography}

\usepackage{graphicx, float, hyperref, tabularx, physics, ulem, multirow,lipsum,supertabular,longtable}	
\usepackage{amsmath}	
\usepackage{amssymb}	

\usepackage{threeparttable} 

\usepackage{caption}
\newcommand{\gx}{{GX~339$-$4}}
\newcommand{\grs}{{GRS~1915+105}}
\newcommand{\exo}{{EXO~1846$-$031}}

\newcommand{\nicer}{{\textit{NICER}}}
\newcommand{\vk}{{\texttt{vKompthdk}}}

\title[Corona seen by \nicer]
{A \nicer\ view of the corona through time-dependent Comptonization of the quasi-periodic oscillations in nine black-hole X-ray binaries}

\author[Y.\ Zhang et al.]{
Yuexin Zhang,$^{1,2}$\thanks{E-mail: yuexin.zhang@cfa.harvard.edu}
Mariano M\'{e}ndez,$^{2}$
James F.\ Steiner,$^{1}$
Federico Garc\'{i}a,$^{3,4}$
Candela Bellavita,$^{2,3,4}$
\newauthor
Ole K\"{o}nig,$^{1}$
Erin Kara,$^{5}$
Santiago Ubach Ramirez,$^{1}$
Honghui Liu,$^{6}$
Zuobin Zhang,$^{7}$
Liang Zhang,$^{8}$
\newauthor
Zi-Xu Yang,$^{9}$
Kevin Alabarta,$^{10}$
Sandeep K.\ Rout,$^{10}$
and Diego Altamirano$^{11}$\\
\\
$^{1}$ Center for Astrophysics \textbar\ Harvard \& Smithsonian, 60 Garden St, Cambridge, MA 02138, USA\\
$^{2}$ Kapteyn Astronomical Institute, University of Groningen, P.O.\ BOX 800, 9700 AV Groningen, The Netherlands\\
$^{3}$ Instituto Argentino de Radioastronom\'{i}a (CCT La Plata, CONICET; CICPBA; UNLP), C.C.5, 1894 Villa Elisa, Argentina\\
$^{4}$ Facultad de Ciencias Astronómicas y Geofísicas, Universidad Nacional de La Plata, 1900 La Plata, Argentina\\
$^{5}$ MIT Kavli Institute for Astrophysics and Space Research, Massachusetts Institute of Technology, Cambridge, MA 02139, USA\\
$^{6}$ Institut f\"ur Astronomie und Astrophysik, Universit\"at T\"ubingen, Sand 1, D-72076 T\"ubingen, Germany\\
$^{7}$ Astrophysics, Department of Physics, University of Oxford, Keble Road, Oxford OX1 3RH, UK\\
$^{8}$ Key Laboratory of Particle Astrophysics, Institute of High Energy Physics, Chinese Academy of Sciences, Beijing 100049, China\\
$^{9}$ School of Physics and Optoelectronic Engineering, Shandong University of Technology, Zibo 255000, China\\
$^{10}$ Center for Astrophysics and Space Science (CASS), New York University Abu Dhabi, P.O.\ Box 129188, Abu Dhabi, UAE\\
$^{11}$ School of Physics and Astronomy, University of Southampton, Southampton, SO17 1BJ, UK\\
}

\date{Accepted XXX. Received YYY; in original form ZZZ}

\begin{document}
\label{firstpage}
\pagerange{\pageref{firstpage}--\pageref{lastpage}}
\maketitle

\begin{abstract}
We present a systematic study of the evolution of the corona geometry in nine black hole X-ray binaries (BHXRBs) using archival data from \textit{NICER}. We identify 171 observations exhibiting quasi-periodic oscillations (QPOs) across various spectral states and model the time-averaged energy spectra of the source, as well as the energy-dependent rms and phase-lag spectra of the QPO, with the time-dependent Comptonization model \texttt{vKompthdk}. This allows us to simultaneously constrain the corona size and feedback fraction during outbursts. By using the power color hue diagnostics, we identify different spectral states, and observe that the QPO frequency increases from $\sim$0.1~Hz to $\sim$10~Hz in the low-hard and hard-intermediate states (LHS and HIMS), and remains approximately constant at 4--5~Hz in the soft-intermediate state (SIMS). The corona size shows significant evolution: the corona is large ($\sim10^4$--$10^5$~km) in the LHS, contracts rapidly to $\sim10^3$~km in the HIMS, and exhibits a flare-like expansion near the HIMS-to-SIMS transition. In the SIMS and high-soft state (HSS), the corona becomes compact and stable (4000--8000~km). The feedback fraction of the corona photons increases during the periods in which the corona contracts and decreases during the periods in which the corona expands, indicating a change of the disk-corona coupling. Our results are consistent with previous QPO-based studies using \vk\ on some individual sources. This work, however, provides the first view of the coronal evolution across outbursts for a diverse BHXRB sample, offering critical insights into coronal behavior as a function of the spectral state of the source.
\end{abstract}




\section{Introduction}\label{sec:intro}

A black hole X-ray binary (BHXRB) system harbors a central BH and a companion star. In the X-ray outburst of such a system, intense X-rays in the soft and hard X-ray bands are observable. The soft X-ray emission is generally from the thermal accretion disk~\citep{1989ESASP.296....3T}, with the disk spectrum peaking at around 0.2--2.0~keV; a fraction of the soft disk photons are Compton up-scattered in the corona, forming a component in the energy spectrum with a shape of a cutoff power-law up to 100~keV~\citep[for reviews, see][]{2007A&ARv..15....1D,2010LNP...794...17G}. In the scenario where a fraction of the Comptonized photons illuminate the accretion disk, these photons are reprocessed and Compton back-scattered, leading to a relativistic reflection component~\citep{1989MNRAS.238..729F}. This reflection spectrum includes characteristic broadened X-ray emission lines, the most prominent of which is the iron $K_{\alpha}$ line at 6.4--7.0~keV, along with the Compton hump around 20~keV~\citep[see][for a review]{2021SSRv..217...65B}.

During a typical outburst, a BH X-ray transient follows an anticlockwise `q' path in the hardness-intensity diagram~\citep[HID;][]{2001ApJS..132..377H,2005Ap&SS.300..107H,2005A&A...440..207B}, as the proportion of the disk and corona components varies, resulting in several well-defined spectral states~\citep{2005A&A...440..207B}: At the onset of the outburst, the source arises from quiescence to the low-hard state (LHS), characterized by a dominant hard corona component and a relatively weak thermal disk component in the X-ray spectrum~\citep[e.g.][]{2015ApJ...813...84G,2018MNRAS.481.5560S}. As the mass accretion rate increases, the source becomes brighter and rapidly transitions through the hard-intermediate state (HIMS), the soft-intermediate state (SIMS), and reaches the high-soft state (HSS)~\citep[e.g.][]{1997ApJ...479..926M,2005A&A...440..207B,2010LNP...794...53B,2012MNRAS.427..595M}. From the LHS to the HSS the truncated disk gradually moves closer to the innermost stable circular orbit (ISCO) around the BH (\citealt{1997ApJ...489..865E}; but see, e.g.~\citealt{2008MNRAS.387.1489R} and~\citealt{2018ApJ...860L..28M} who argue that the disk is not truncated in the LHS/HIMS) and the thermal component becomes dominant, while the spectrum of the hard component becomes steeper and weaker~\citep[e.g.][]{2018MNRAS.481.5560S,2022MNRAS.514.1422D}. In the decay of the outburst, with a gradual decay of the mass accretion rate, the source transitions back to the LHS at a lower luminosity than the forward transition before finally returning to the quiescent state~\citep{2021MNRAS.507.5507A}.

In addition to X-ray emission during the outburst, radio emission from a jet is often prominent~\citep[for a review, see][]{2006csxs.book..381F}. The relationship between the spectral states and the radio emission suggests the existence of a corona-jet coupling~\citep{2022NatAs...6..577M}. In both the LHS and HIMS, a compact jet is observed; however, during the state transition from the HIMS to the SIMS, the compact, optically thick, jet is usually quenched~\citep{2009MNRAS.396.1370F}. In the SIMS, the compact jet is no longer present, but a bright transient jet appears, accompanied by observable discrete, optically thin, relativistic ejecta~\citep{1994Natur.371...46M}. Later in the HSS, the jet disappears~\citep{2004ApJ...617.1272C}.

Open questions persist regarding the corona geometry near BHs and its evolution during an outburst. Spectral analyses indicate the corona is located within a truncated disk that approaches the ISCO from the LHS to the HSS during the outburst~\citep{2007A&ARv..15....1D,2013MNRAS.430.3196V}. This scenario has sometimes been confronted by a non-truncated disk picture with a lamppost corona structure located above the BH, which has been supported by the reflection modeling of relativistically-broadened iron lines in the LHS~\citep[e.g.][]{2018ApJ...852L..34X}. However, the current polarimetric observations suggests a corona extended in the disk plane, posing significant questions on the current understanding of the corona geometry inferred from reflection spectroscopy fitting~\citep{2022Sci...378..650K,2024ApJ...968...76I,2024ApJ...969L..30S}. The conflicting spectral and polarimetric results demonstrates an urgent need for new evidence to constrain the disk-corona geometry across state transitions.

X-ray timing analysis is a powerful tool for exploring the innermost regions of BHs in XRBs~\citep[see][for a recent review]{2019NewAR..8501524I}. In the Fourier domain, the power density spectrum (PDS) of the X-ray light curve shows different kinds of variability, e.g., broadband noise and narrow peaks called quasi-periodic oscillations~\citep[QPOs;][]{1989ASIC..262...27V,2000MNRAS.318..361N,2002ApJ...572..392B}. The most frequent QPOs in BHXRBs are the low-frequency QPOs (LFQPOs), with central frequencies ranging from mHz to $\sim$ 30~Hz~\citep{2014SSRv..183...43B}. These QPOs are classified into types A, B, and C, depending on the shape of the broadband noise, the broadband fractional root-mean-square amplitude (hereafter rms), the phase lags of the QPOs, and the spectral state of the source~\citep{2002ApJ...564..962R,2004A&A...426..587C,2005ApJ...629..403C}. Type-C QPOs generally appear in the LHS and HIMS, type-B QPOs only appear in the SIMS, and type-A QPOs, which are weak and rarely detected, appear in the SIMS and HSS~\citep{2016ASSL..440...61B}. In the short-lived SIMS, fast QPO transitions are sometimes observed between the type-B and either other types of QPOs or the disappearance of QPOs~\citep[e.g.,][]{2004A&A...426..587C,2021MNRAS.505.3823Z,2023MNRAS.521.3570Y}.

The Fourier cross spectrum of two simultaneous light curves in different energy bands (subject and reference band) can be used to compute the phase lags as a function of Fourier frequency, probing the physics of the innermost region near a BH~\citep[see][for a review]{2014A&ARv..22...72U}. Traditionally, it is common to report lags of the signals, e.g., the broadband noise and the QPOs, identified in the PDS extending over a given characteristic frequency range~\citep{1999ApJ...510..874N}. Recently, a novel method has been proposed to compute the phase lag based on Lorentzian function fitting to the real and imaginary parts of the cross spectrum~\citep{2024MNRAS.527.9405M}. This method has an advantage over the traditional method for measuring the phase lag when the Fourier components are not dominant over a given frequency range in the PDS. For the measured phase lags, hard (positive) lags~\citep{1988Natur.336..450M} can be produced by the propagation of fluctuations in the mass accretion rate from the outer part towards the inner part of the disk and the corona~\citep[e.g.,][]{2006MNRAS.367..801A,2013MNRAS.434.1476I}. Another explanation for the hard lags is the Comptonization in the jet~\citep{2008A&A...489..481K,2024A&A...690A...6K}. Soft (negative) lags can arise due to reverberation or thermalization of hard photons when the corona photons impinge back onto the accretion disk~\citep[e.g.,][]{2009Natur.459..540F,2020MNRAS.492.1399K}.

Various models have been proposed to explain the time variability signals from the corona and constrain the corona geometry in BHXRBs. These include the model for reverberation lags and lamppost/cylinder geometry~\citep{2019MNRAS.488..324I,2023ApJ...951...19L}, the corona outflow model for wind-like structures~\citep{2008A&A...489..481K}, the propagation of mass accretion rate fluctuations model for truncated disks~\citep[e.g.][]{2013MNRAS.434.1476I,2016MNRAS.462.4078R,2025MNRAS.536.3284U}, the JED-SAD model for jet-emitting and standard accretion disks~\citep{1997A&A...319..340F,2018A&A...615A..57M}, and the Lense-Thirring precession or disk instability models for explaining LFQPOs and hard-soft lags~\citep{1998ApJ...492L..59S,2009MNRAS.397L.101I}. 

One of the most compelling methods to measure the corona geometry is to jointly model the time-averaged energy spectrum of the source, together with the energy-dependent rms and phase-lag spectra at the QPO frequency, using the time-dependent Comptonization model \vk~\citep{2020MNRAS.492.1399K,2022MNRAS.515.2099B}. This model considers the physical process of photons that can be both direct and inverse-Comptonized in the corona and thermally reprocessed in the disk. In addition to the temperatures of the disk and corona, some critical parameters in \vk, i.e., the corona size, $L$, and the corresponding feedback fraction, $\eta$, constrain the physical geometry of the corona. The feedback fraction, $\eta$, ranging between 0 and 1, quantifies the thermal flux produced by hard photons reprocessed by the disk relative to the observed disk flux. Key insights about these critical parameters are: If $\eta$ is high or close to 1, the corona extends primarily within the disk plane; If $\eta$ is low or close to 0, the corona is extended perpendicularly to the disk or expanding away from the disk~\citep[e.g.][]{2022NatAs...6..577M}. In practice, $\eta$ is often converted into the intrinsic feedback fraction, $\eta_{\text{int}}$~\citep{2020MNRAS.492.1399K}, which represents the corona's feedback efficiency~\citep{2023MNRAS.520.5144Z,2023MNRAS.525..854M,2025ApJ...980..251A}. It quantifies the fraction of Comptonized photons that return to the disk relative to the total flux produced by Comptonization.

By combining measurements of corona size and feedback fraction, we can better understand the geometry of the corona and its evolution. Past studies have successfully applied \vk\ to fit the rms and phase-lag spectra of the QPO, and the time-averaged energy spectra of the source for seven BHXRBs~\citep{2021MNRAS.501.3173G,2022NatAs...6..577M,2022MNRAS.512.2686Z,2022MNRAS.513.4196G,2023MNRAS.519.1336P,2023MNRAS.520..113R,2023MNRAS.520.5144Z,2023MNRAS.525..221R,2023MNRAS.525..854M,2023MNRAS.526.3944Z,2025ApJ...980..251A,2025A&A...697A.229R}, among which only four sources have been studied using \nicer\ data, and only one has been investigated throughout the outburst. These results reveal how the corona extends over the disk and how its physical geometry evolves during (part of) the spectral state transitions of BHXRBs. 

In this paper, we explore the \nicer\ archival data of bright BH XRBs~\citep{2022ApJ...930...18W} across the spectral state transitions during outbursts. We present the evolution of the corona properties of nine sources with QPOs using \vk~\citep{2020MNRAS.492.1399K,2022MNRAS.515.2099B}. In Section~\ref{sec:analysis}, we describe the observations and data analysis. In Section~\ref{sec:results}, we show the evolution of the time variability and corona geometry during different stages of the state transitions. Finally, in Section~\ref{sec:discussion}, we compare the coronal properties from this systematic study with our previous studies on individual sources. Additionally, we discuss the corona evolution from modeling QPO properties and other approaches, e.g., reverberation.

\section{Observations and data analysis}\label{sec:analysis}

Our investigation begins with the ten bright BHXRBs reported by \citet{2022ApJ...930...18W} to ensure robust measurements of time variability in these sources, plus new bright BHXRBs that reach a peak rate > 100~mCrab. We use the standard \texttt{nicerl2} (NICERDAS v12) routine to reduce the data. Noisy detectors \#14 and \#34 are always excluded~\footnote{\url{https://heasarc.gsfc.nasa.gov/docs/nicer/data_analysis/nicer_analysis_tips.html\#noisy_detectors}}. We also exclude the seven detectors in MPU1 due to the time stamp anomalies from July 8--23 2019~\footnote{\url{https://heasarc.gsfc.nasa.gov/docs/nicer/data_analysis/nicer_analysis_tips.html\#July2019-MPU1_Timing_Errors}}. To avoid interruptions caused by telemetry saturation, which can result in a large number of short Good Time Intervals (GTIs) with brief gaps between them, we use only continuous GTIs of minimal time of 17~s. Such short gaps between consecutive GTIs can introduce spurious time signatures in the PDS, irrelevant to the X-ray source of interest. Additionally, after May 2023 only orbital night data are considered to avoid the effects of optical leaking\footnote{\url{https://heasarc.gsfc.nasa.gov/docs/nicer/analysis_threads/light-leak-overview/}}.

\subsection{Power density spectrum}\label{subsection:pds}

We generate the PDS for each observation and for each \nicer\ orbit within an observation in the 0.3--12~keV energy band, as well as in six narrow bands: 0.3--1.8, 1.8--3, 3--4, 4--5.5, 5.5--8, 8--12~keV. The length of each PDS segment is 131.072~s, and the Nyquist frequency is 250~Hz. In each observation, we average all the PDS segments, subtract the Poisson noise using the average power in the frequency range $120$--$250~\text{Hz}$, and normalize the PDS to fractional rms amplitude~\citep{1990A&A...230..103B}. Finally, we apply a logarithmic rebinning in frequency to the PDS, such that the bin size increases by a factor of $10^{1/100}\approx 1.023$ compared to the previous bin.

We fit the averaged PDS of each observation in the $0.3$--$12~\text{keV}$ band using a linear combination of Lorentzian functions~\citep{2000MNRAS.318..361N,2002ApJ...572..392B} over the frequency range $\sim$$0.03$--$30~\text{Hz}$. The parameters of a Lorentzian function are its central frequency ($\nu_0$), full width at half maximum (FWHM), and normalization. 
All PDS are modeled with five Lorentzians, representing the QPO fundamental, the second harmonic, the sub harmonic, and two broadband noise components. Additional Lorentzians are included if extra broadband features are present. All parameters of the Lorentzians are left free during fitting, except for the central frequency of one Lorentzian function, which is fixed at $0$ to model the zero-centered broadband noise. In cases where no QPO is detected, the number of Lorentzian functions is reduced to one or two to fit only the broadband noise.
After the initial fitting, if the QPO frequency of an observation exhibits significant changes with time, indicating a fast transition nature of the QPO, we divide the observation into separate \textit{NICER} orbits. This ensures that the QPO feature is well-modeled by a single Lorentzian.

Using the models that we apply to fit the PDS of each observation or orbit in the 0.3--12~keV band as baselines, we further fit the PDS of each orbit in the six narrow energy bands (see above). We check that the central frequency and width of those Lorentzians do not change significantly with energy. We subsequently fix the central frequencies and FWHMs of all the Lorentzians to the values we obtained for the PDS in the 0.3--12~keV band and only let the normalizations free to fit the PDS in those six narrow energy bands. Finally we take the square root of the normalizations of the Lorentzians that represent the variability components to calculate the fractional rms. We note that the energy-dependent rms spectra presented in this work specifically refer to the QPO component, which is treated as a single, coherent variability mode, rather than a superposition of independently varying spectral components.

To build our sample of QPO data, we require that the QPO has a quality factor ($Q = \nu / \text{FWHM}$) larger than 2~\citep[e.g.][]{1995xrbi.nasa..252V,2002ApJ...572..392B,2012MNRAS.423..694R,2018ApJ...865L..15S} and is dominant over the underlying broadband noise. 
Additionally, we require that the fractional rms amplitude and phase lag, the latter of which also depends on the QPOs fitted by Lorentzians~\citep{2024MNRAS.527.9405M}, are well constrained across all five or six narrow energy bands. In practice, observations are excluded if the fractional rms of the QPO in the full energy band is consistent with zero within $3\sigma$ uncertainties, or if the fractional rms of the QPO cannot be reliably measured (i.e.\ $<1\sigma$) in one or more individual energy bands.
These criteria rule out some observations with QPOs, e.g.\ those of MAXI~J1820+070 prior to observation ID~1200120138~\citep{2025MNRAS.542..350B}. Finally, 171 \textit{NICER} observations or orbits (hereafter referred simply as observations) from nine sources are selected and reported in Table~\ref{tab:obs}. Note that for the sources EXO~1846$-$031~\citep{2021ApJ...906...11W}, GRS~1915+105~\citep{2018ApJ...860L..19N}, MAXI~J1535$-$571~\citep{2022MNRAS.512.2686Z}, and MAXI~J1631$-$479~\citep{2021MNRAS.505.1213R}, the Galactic absorption is high, $> 10^{22}~\text{cm}^{-2}$, so we exclude the lowest energy band for the rms and phase-lag (see below) calculation in these cases.

\subsection{Power color}

Power color (PC) is an efficient way to identify the spectral state across different BHXRBs~\citep{2015MNRAS.448.3339H}. It characterizes the shape of the PDS independently of its normalization, in a manner analogous to how spectral hardness characterizes the shape of the flux-energy spectrum. We define PC1 as the integrated power between $0.25$ and $2.0~\text{Hz}$ divided by that between $0.0039~\text{Hz}$ and $0.031~\text{Hz}$. Similarly, PC2 is defined as the integrated power between $0.031~\text{Hz}$ and $0.25~\text{Hz}$ divided by that between $2.0~\text{Hz}$ and $16.0~\text{Hz}$, with the PDS computed in the $4.8$--$9.6~\text{keV}$ band~\citep{2015MNRAS.448.3339H,2022ApJ...930...18W}.

In a PC diagram (PCD), we define the origin at PC1 $= 4.5$ and PC2 $= 0.45$ to characterize spectral transitions using the angle ``hue'' in the PC wheel~\citep{2015MNRAS.448.3339H,2022ApJ...930...18W}. Specifically, the hue is defined as the clockwise angle from the axis oriented at $45^{\circ}$ to the $-x$ and $+y$ axes. The hue ranges corresponding to different spectral states are as follows: the LHS spans from $-20^{\circ}$ to $140^{\circ}$, the HIMS from $140^{\circ}$ to $220^{\circ}$, the SIMS from $220^{\circ}$ to $300^{\circ}$, and the HSS from $300^{\circ}$ to $-20^{\circ}$~\citep{2015MNRAS.448.3339H}.

\subsection{Phase lag spectrum}\label{subsec:phase lag}

We compute fast Fourier transforms (FFTs) using data segments of 32.768~s duration and a Nyquist frequency of 125~Hz for each observation. The FFTs are calculated in the 0.3--12~keV band, which we use as the reference band, and the same six narrow energy bands as in Section~\ref{subsection:pds}, which we use as the subject bands. For each segment, we compute the cross spectrum between the reference and each subject band. 
The real and imaginary parts of the cross spectrum are averaged over all segments within each observation. The averaged real part in the 70--125~Hz, where the source does not contribute to the variability, is then subtracted from the real part of the cross spectrum. This step removes the partial correlation between the subject and reference bands, which arises because the subject band is included in the reference band~\citep{2019MNRAS.488..324I,2024MNRAS.527.9405M,2024MNRAS.527.7136B}.
The lowest energy band is excluded for some sources, as we do for the PDS.
The phase lag is $\phi(f) = \arctan \left(\text{Im}\,G(f)/\text{Re}\,G(f)\right)$, where the cross spectrum $G$ is a complex function of Fourier frequencies. Thus, we can calculate the phase lags of the variability components measured in the PDS~\citep{1999ApJ...510..874N,2019MNRAS.489.3927I}.

As explained in~\citet{2024MNRAS.527.9405M}, we obtain phase lags across different energy bands by simultaneously fitting both the real and imaginary parts of $G$ using the same Lorentzian components that are used to fit the PDS~\citep{2022MNRAS.513.2804P,2022MNRAS.514.2839A,2023MNRAS.520.5144Z,2025ApJ...980..251A}. 
For the real and imaginary parts, we fit the normalizations of the Lorentzian components, where the phase lag, $\phi$, for each Lorentzian is given by the arctangent of the ratio of the normalization in the imaginary part to that in the real part. This approach assumes that the phase lag of each component remains constant with frequency~\citep{2022MNRAS.513.2804P,2024MNRAS.527.9405M}. 
Using this method, we compute the phase lags of all variability components fitted by the Lorentzian functions for the defined energy bands.

\subsection{Energy spectral analysis}

We use the \texttt{nicerl3-spec} tool to extract the source and the \texttt{3C50} background spectra~\citep{2022AJ....163..130R} in each observation. We group the spectra using optimal binning~\citep{2016A&A...587A.151K}, with the additional requirement of 10 counts per grouped bin.
We fit the energy spectra using XSPEC v12.14.0~\citep{1996ASPC..101...17A}.

We apply the baseline model \texttt{TBfeo*(diskbb+vKompthdk+gaussian)} to fit the energy spectra. The component \texttt{TBfeo} models the Galactic absorption between the source and observer with variable oxygen and iron abundances. We set the cross section and the solar abundance of the interstellar medium (ISM) using the tables of~\citet{1996ApJ...465..487V} and~\citet{2000ApJ...542..914W}, respectively. 
The component \texttt{diskbb}~\citep{1984PASJ...36..741M} represents the emission of a multi-temperature, optically thick, and geometrically thin disk, with its parameters being the disk temperature, $kT_{\text{in}}$, and a normalization factor. 
The time-averaged (steady-state) version of the time-dependent Comptonization model \texttt{vKompthdk}~\citep{2020MNRAS.492.1399K,2022MNRAS.515.2099B} is equivalent to \texttt{nthcomp}~\citep{1996MNRAS.283..193Z,1999MNRAS.309..561Z}. The parameters of this component are the seed photon temperature, $kT_{\text{s}}$, the corona temperature, $kT_{\text{e}}$, the photon index, $\Gamma$, and a normalization factor. The coronal electron temperature, $kT_{\text{e}}$, is fixed at 50~keV, since the \nicer\ energy band does not provide sufficient constraints. The seed photon temperature, $kT_{\text{s}}$, in \texttt{vKompthdk} is linked to the inner disk temperature, $kT_{\text{in}}$, in \texttt{diskbb}. Finally, we add a Gaussian component, with its central energy constrained to be in the 6.4--7~keV energy range, to fit the broad iron $K_{\alpha}$ emission line in the spectrum. 
For observations significantly affected by \nicer\ calibration systematics, the most noticeable features include edge-like structures near $\sim$0.5keV (oxygen), $\sim$1.8keV (aluminum K), and $\sim$2.3keV (gold M). We model these features empirically using \texttt{edge} components (see summary in Table~\ref{tab:obs}).

Note that the \texttt{vKompthdk} model contains four additional parameters that only affect the time-dependent spectrum, describing the corona size, $L$, the feedback fraction, $\eta$, the amplitude of the variability of the rate at which the corona is heated by an (unspecified) external source, $\delta\dot{H}_{\text{ext}}$, and an additive parameter, \texttt{reflag}, that gives the phase lag in the 2--3~keV band. The parameter \texttt{reflag} provides an additive offset to the lag-energy spectrum, anchoring the model to the observed lag in the 2--3~keV band, while the other parameters define the shape of the phase-lag spectrum. None of these four parameters alter the steady-state spectrum produced by \texttt{vKompthdk}. These four parameters, along with $kT_{\text{s}}$, $kT_{\text{e}}$, and $\Gamma$, describe the radiative properties of the QPOs, namely the rms and the phase lags. 

The spectral fitting yields acceptable results for most sources. An exception is \grs, for which several observations exhibit spectral features associated with X-ray winds. These winds are indicated by the presence of narrow, blue-shifted absorption lines from highly ionized Fe~XXV ($\sim$6.7~keV) and Fe~XXVI ($\sim$6.97~keV). To account for these features and improve the fit, we include gaussian absorption components to model the absorption lines.

\subsection{Joint fitting}

In the previous subsection, we described how we have computed the rms and phase-lag spectra of QPOs and separately fitted the time-averaged energy spectrum of the source for each observation. Using these spectra and the initial parameter values, we subsequently perform a joint fit of the rms and phase-lag spectra of the QPO, along with the energy spectrum of the source, for each of the 171 observations (see samples in Appendix~\ref{sec:appendix samples}). The model we use is \texttt{TBfeo*(diskbb+vKompthdk+gaussian)}, where \texttt{vKompthdk} includes a mode that enables simultaneous fitting of the rms and phase-lag spectra. For \grs, the model additionally includes gaussian absorption components. 
The \vk\ model assumes that the QPO represents a small oscillation around the steady-state solution of the Kompaneets equation~\citep{1957JETP....4..730K}, which describes the evolution of the photon energy distribution through Comptonization in the isotropical corona. As such, it provides a physical framework that connects the energy-dependent rms and phase lags with the thermal and geometric properties of the disk-corona system.

Note that for the fitting of the rms spectrum of the QPO, \texttt{vKompthdk} is multiplied by a \texttt{dilution} component, which is not explicitly included in the total model, and which does not introduce any new parameters to the fits~\citep{2023MNRAS.520.5144Z,2023MNRAS.525..854M,2025ApJ...980..251A,2025A&A...697A.229R}. The introduction of the \texttt{dilution} component is justified since the time-dependent Comptonization model computes the rms spectrum of a variable corona, whereas the observed rms is reduced by all non-variable spectral components. Since we isolate the QPO contribution in the cross spectrum and model its lags within \texttt{vKompthdk}, the measured phase lags are not diluted by other spectral components, such as the disk, which is not assumed to vary independently on the QPO timescale. Here, we assume that the \texttt{diskbb} and \texttt{gaussian} components are not variable~\citep{2001MNRAS.321..759C}. Therefore, the \texttt{dilution} component is given by
\begin{equation}
   \text{dilution} = \frac{\text{Flux}_{\text{Compt}}(E)}{\text{Flux}_{\text{Total}}(E)} 
\end{equation}
such that
\begin{equation}
\text{rms}_{\text{Obs}} = \text{rms}_{\text{Compt}} \times \frac{\text{Flux}_{\text{Compt}}}{\text{Flux}_{\text{Total}}}.
\end{equation}

When we only fit the energy spectrum, we link $kT_{\text{s}}$ in \texttt{vKompthdk} to $kT_{\text{in}}$ in \texttt{diskbb}. However, when fitting simultaneously the rms and phase-lag spectra of the QPO and the time-averaged energy spectrum of the source, we allow $kT_{\text{s}}$ to vary freely.
Based on our analysis, letting $kT_{\text{s}}$ vary freely provides a better fit (See also~\citealt{2023MNRAS.520..113R,2023MNRAS.520.5144Z,2023MNRAS.525..854M,2025ApJ...980..251A} where we did the same).

We note that, although a fit with a dual corona~\citep[see][]{2021MNRAS.501.3173G} may further reduce the $\chi^{2}$, we do not explore this approach here due to the limited number of energy bands in the rms and phase-lag spectra of the QPO, as well as the large number of free parameters that would be involved.

\section{Results}\label{sec:results}

\subsection{Power color diagram}\label{subsec:pc}

\begin{figure*}
    \includegraphics[width=0.9\textwidth]{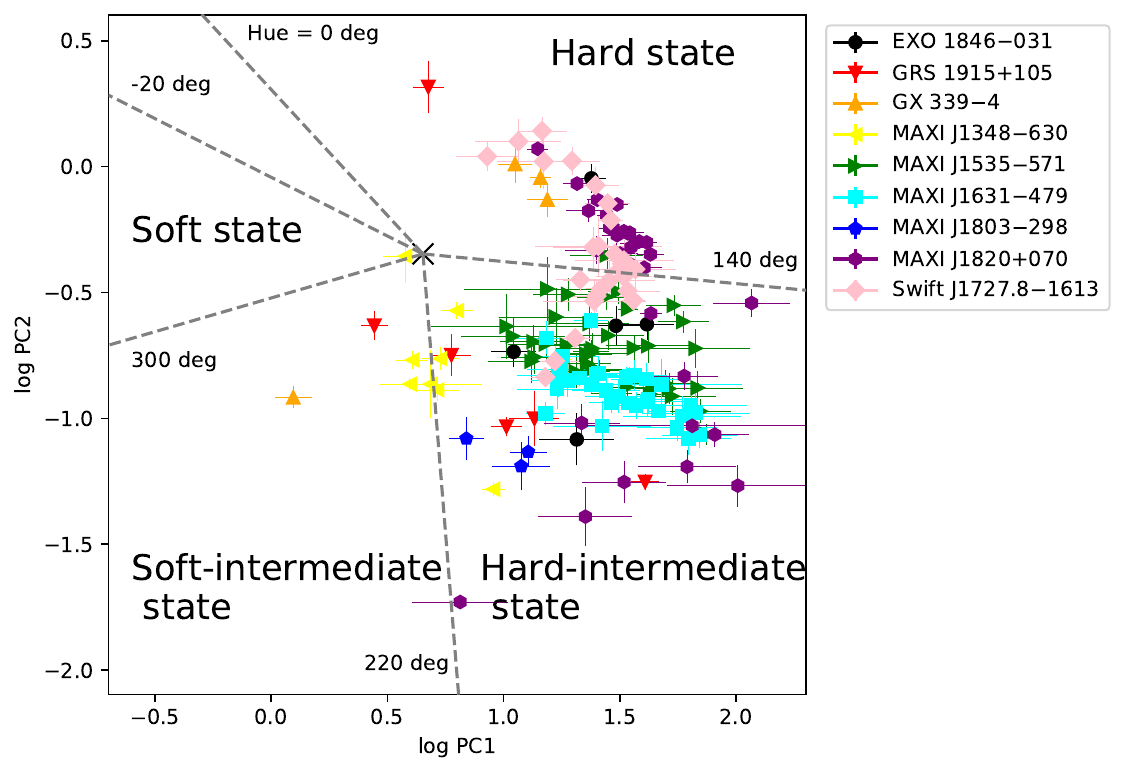}
    \caption{Power-color diagram (PCD) for observations with QPOs from nine selected BHXRBs. The $x$-axis ($\log \text{PC1}$) and $y$-axis ($\log \text{PC2}$) represent the two PC components derived from the PDS. Different colors and markers correspond to different sources, as indicated in the legend. The diagram is divided into four spectral states: the LHS, HIMS, SIMS, and HSS, with boundaries marked by grey dashed lines. The hue, representing the angular position in the PC1-PC2 plane, is labeled at specific angles. The distribution of points shows how different sources evolve through various spectral states. Error bars indicate 1-$\sigma$ uncertainties.}
    \label{fig:pcd}
\end{figure*}

In Figure~\ref{fig:pcd}, we show the PCD divided into the four primary spectral states: the LHS, HIMS, SIMS, and HSS. These divisions follow previous studies on the spectral evolution of BHXRBs~\citep{1997ApJ...479..926M,2005A&A...440..207B,2022ApJ...930...18W}, with the boundaries marked by dashed lines in the plot. The PC hue is also annotated at specific positions, namely, $0^\circ$ for the reference point and $140^\circ$, $220^\circ$, and $300^\circ$ for the transitions between states. 

The data points are only from observations with QPO detections.
Observations in the LHS cluster around high $\log \text{PC1}$ and \(\log \text{PC2}\) values, indicating that the PDS is dominated by low-frequency variability. Although we only detect one QPO in the HSS, where the error bar of the PC is consistent with being in the SIMS, \citet{2015MNRAS.448.3339H} report that this state is characterized by low $\log \text{PC1}$ and $\log \text{PC2}$ values, where variability is significantly weaker. The HIMS and SIMS represent the state transition between the hard and soft states. The PCD effectively captures the spectral evolution of BHXRBs and the hue will be further used to study the connection between the time variability properties and corona geometry.

We summarize the QPO frequency and hue for each of the observations in Table~\ref{tab:obs}. The behaviors described below are those associated with the presence of QPOs. According to the PCD, \grs\ is seen in the LHS, HIMS and SIMS. For \gx, and MAXI~J1820+070, the state transitions occur from the LHS to the HIMS, followed by the SIMS. 
\exo, MAXI~J1535$-$571, and Swift~J1727.8$-$1613 undergo transitions from the LHS into the HIMS but show no further indication of transitioning into the SIMS or HSS. 
Only MAXI~J1631$-$479 and MAXI~J1803$-$298 are located in the HIMS, lacking clear state transition behavior. 
MAXI~J1348$-$630 exhibits a transition from the HIMS to the SIMS and marginally into the HSS.

\subsection{QPO and hue evolution}\label{subsec:qpo}

\begin{figure*}
    \includegraphics[width=0.9\textwidth]{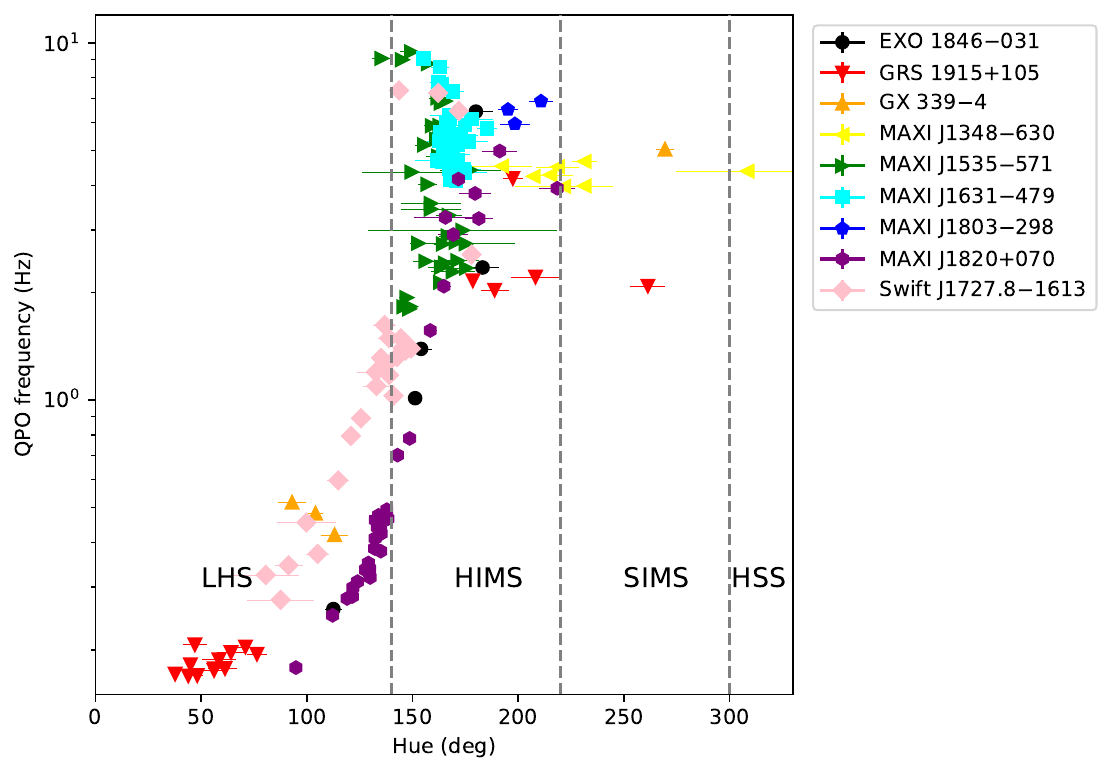}
    \caption{QPO frequency as a function of power color hue for 9 selected BHXRBs. Different colors and symbols represent different sources, as indicated in the legend. The vertical dashed lines separate the LHS, HIMS, SIMS, and HSS. The QPO frequency increases systematically with hue, reflecting the spectral state evolution of BHXRBs. Error bars indicate 1-$\sigma$ uncertainties.}
    \label{fig:qpo_hue}
\end{figure*}

Figure~\ref{fig:qpo_hue} presents the relationship between QPO frequency and PC hue during outbursts across different spectral states of nine selected BHXRBs. Each data point corresponds to an individual observation, with different colors and symbols representing different sources. The dashed grey lines are anchored at $140^\circ$, $220^\circ$, and $300^\circ$, corresponding to the LHS-HIMS, HIMS-SIMS, and SIMS-HSS transitions, respectively.

In the LHS and HIMS, the QPO frequency monotonically increases from $\sim$0.1~Hz to 10~Hz as the hue increases, showing a strong correlation with hue. Parallel tracks arise since hue is determined by the integrated power in broad frequency bands, and not by the QPO frequency. Consequently, sources with different QPO frequencies can exhibit the same hue when the total power in the frequency range containing the QPO is comparable.
During the HIMS-SIMS transition, the QPO frequency remains more or less constant around 4--5~Hz. The increasing trend in QPO frequency is consistent with the behavior of type-C QPOs in the LHS and HIMS, while the nearly constant QPO frequency in the SIMS aligns with the characteristics of type-B QPOs~\citep{2019NewAR..8501524I}. It has been shown that QPO frequency can be an indicator of the spectral state~\citep[e.g.,][]{2005A&A...440..207B}. The strong correlation between QPO frequency and hue highlights the effectiveness of the PC method in tracking spectral state transitions.

Five observations of EXO~1846$-$031 (black circles) evolve from the LHS to the HIMS, overlapping with those of MAXI~J1820+070. Their hue values range from $\sim$$110^\circ$ to $180^\circ$. In the LHS, the QPO frequency starts at around 0.2--0.3 Hz and increases as the source transitions toward the HIMS, reaching $\sim$ 6 Hz.

GRS~1915+105 (red downward triangles) remains mostly in the LHS with hues below $100^\circ$, where it exhibits low-frequency QPOs at around 0.2~Hz. Some observations indicate change of the state, with QPO frequencies reaching approximately 3~Hz.

The data of GX~339$-$4 (orange upward triangles) are scattered between the LHS and SIMS, with hue values around $100^\circ$ and $270^\circ$, respectively. Due to the lack of enough QPO measurements, no clear evolution trend of the QPO frequency is observed. Instead, only a few data points show QPOs around 0.5 Hz in the LHS and around a few Hz in the SIMS.

Following a transition path from the HIMS to the SIMS, MAXI~J1348$-$630 (yellow leftward triangles) spans hues from $180^\circ$ to $300^\circ$. The QPO frequency remains approximately at 4.5~Hz as this source moves toward softer states, consistent with a typical type-B QPO during BHXRB outbursts~\citep{2020MNRAS.496.4366B}.

Observations of MAXI~J1535$-$571 (green rightward triangles) show that the QPO is mostly detected in the HIMS, with hues in a narrow range of $\sim$$140^\circ$ to $170^\circ$. No clear correlation between hue and QPO frequency is observed, with QPO frequencies reaching up to 10~Hz in the HIMS.

MAXI~J1631$-$479 (cyan squares) exhibits a narrower distribution of hue and QPO frequency in the HIMS compared to MAXI~J1535$-$571. The hue of this source is around $170^\circ$, while the QPO frequency ranges from 4~Hz to 10~Hz, similar to that of MAXI~J1535$-$571.

In the case of MAXI~J1803$-$298 (blue pentagons), three observations are in the HIMS, where the hue values are around $200^\circ$, close to the transition to the SIMS. These data points occupy the upper-right region of the HIMS (high hue and high QPO frequency), marking the high end of the correlation between QPO frequency and hue.

MAXI J1820+070 (purple hexagons) spans a wide range from the LHS to the edge of the SIMS, with hues varying from $\sim$$100^\circ$ to $220^\circ$. In the LHS, the QPO frequency of this source starts at $\sim$0.2~Hz, increases to 5 Hz in the HIMS, and then slightly decreases to 4.5 Hz in the SIMS. Among all sources, this one exhibits the most well-defined QPO evolution.

Swift J1727.8$-$1613 (pink diamonds) is located in the LHS and HIMS, with hues ranging from $80^\circ$ to $180^\circ$. Its QPO frequency falls within the range of 0.3--7 Hz. Due to Sun-angle constraints, this source lacks observations in the SIMS~\citep{2024ApJ...968...76I}.

\subsection{Ratio of inner-disk to seed-photon temperatures}\label{ratio}

\begin{figure}
    \includegraphics[width=0.45\textwidth]{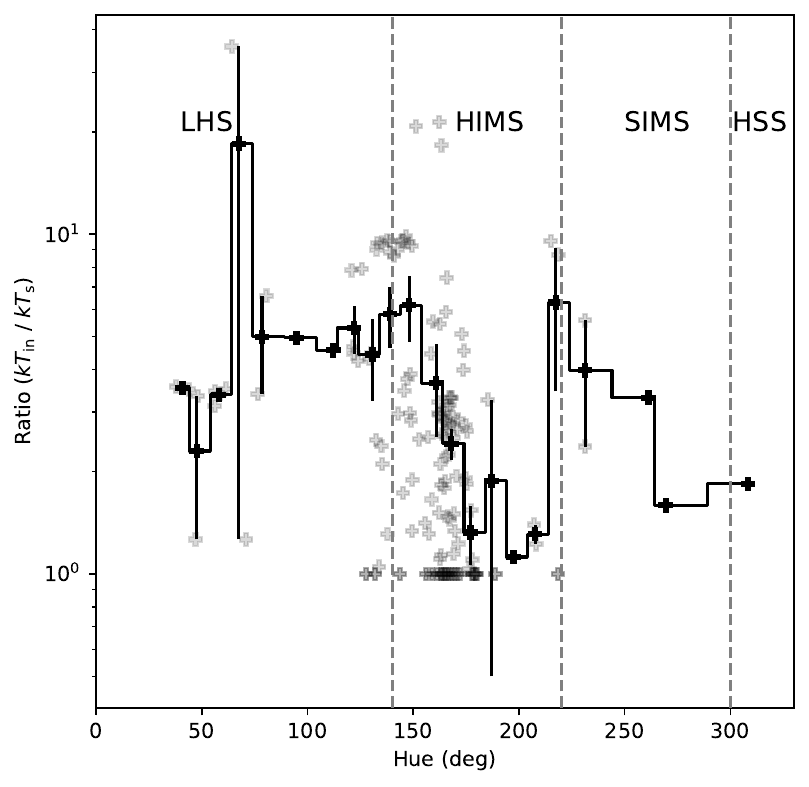}
    \caption{Ratio between the inner disk temperature, $kT_{\rm in}$, of \texttt{diskbb} and the seed photon temperature, $kT_{\rm s}$, of \vk\ as a function of power-color hue. Grey points show individual measurements from observations of nine sources, while black points represent values averaged in bins of $10^\circ$ in hue, with error bars represent the uncertainty on the mean within each bin. Vertical dashed lines mark the boundaries between the LHS, HIMS, SIMS, and HSS.}
    \label{fig:ratio_hue}
\end{figure}

In the joint fitting of the energy spectrum and the QPO rms and phase-lag spectra, we allow the seed photon temperature, $kT_{\rm s}$, in \texttt{vKompthdk} to vary
independently from the inner disk temperature, $kT_{\rm in}$, inferred from the \texttt{diskbb} component. This choice is physically motivated: In \vk\ framework, $kT_{\rm s}$ characterizes the photons entering the variable Comptonization region that produces the QPO, and therefore does not necessarily correspond to the effective inner disc temperature inferred from the time-averaged spectrum~\citep{2022MNRAS.515.2099B}. Allowing $kT_{\rm s}$ to vary independently enables us to probe the physical origin of the seed photons, rather than assuming that they always arise from the same region of the disc traced by $kT_{\rm in}$~\citep[e.g.][]{2023MNRAS.520..113R,2023MNRAS.520.5144Z,2023MNRAS.525..854M}.
To examine whether $kT_{\rm s}$ follows the same evolution as $kT_{\rm in}$ across different spectral states, we show in Figure~\ref{fig:ratio_hue} the ratio $kT_{\rm in}/kT_{\rm s}$ as a function of power-color hue. Grey symbols show individual measurements, while black points represent values averaged in bins of $10^\circ$ in hue.

In the LHS ($\text{hue} < 140^\circ$), the ratio significantly deviates from unity, indicating that the seed photon temperature differs from the inner disk temperature inferred from the time-averaged spectrum. As the source evolves from the $\text{hue}$ of $140^\circ$ to $\sim200^\circ$ toward HIMS, the ratio shows a systematic evolution, gradually approaching unity through the HIMS. From during the HIMS-to-SIMS transition near the hue of $220^\circ$, the ratio increases again. From the SIMS to the HSS ($220^\circ < \text{hue} < 300^\circ$), the ratio decreases and becomes close to unity. This systematic, state-dependent behavior provides an insight on the origin and evolution of the seed photons that are Comptonized in the variable corona, and will be discussed in Section~\ref{discussion:seed photons}.

\subsection{Corona geometry evolution}\label{subsec:corona}

\begin{figure*}
    \includegraphics[width=0.9\textwidth]{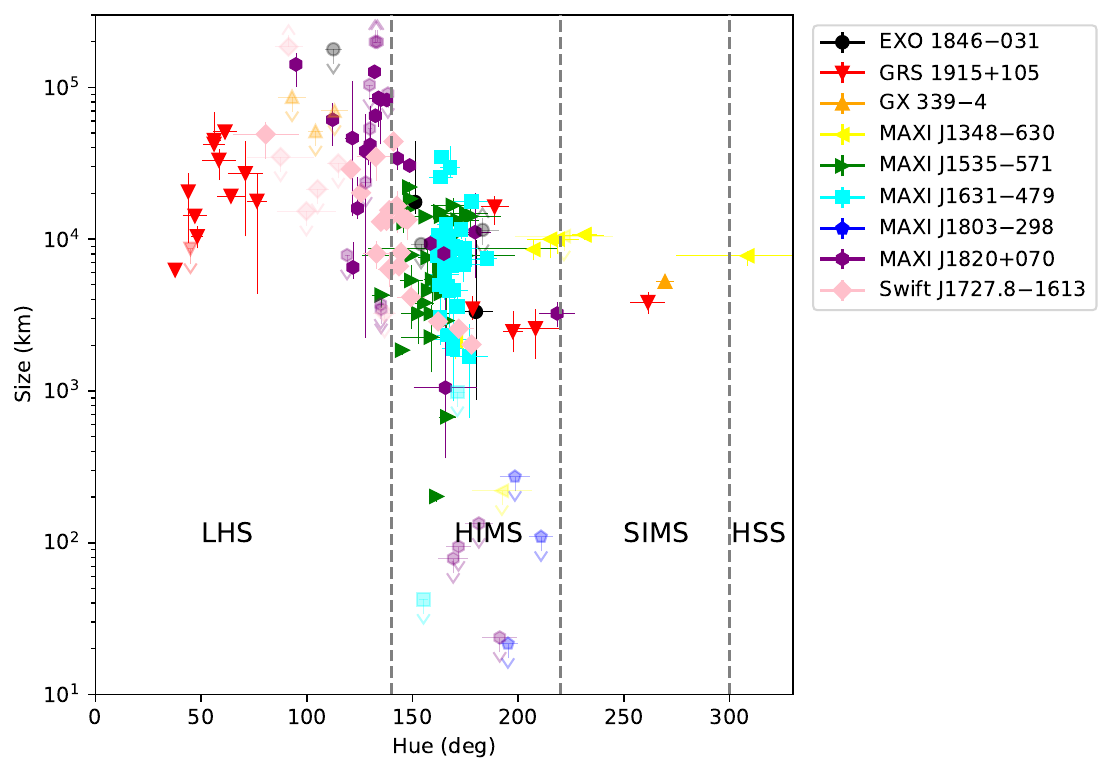}
    \caption{Corona size as a function of PC hue for nine BHXRBs. Different colors and symbols represent different sources, as indicated in the legend. The vertical dashed lines, located at hue values of $140^\circ$, $220^\circ$ and $300^\circ$, mark the boundaries between the LHS, HIMS, SIMS, and HSS. The error bars indicate 1-$\sigma$ uncertainties. The points with arrows indicate the 90\% lower or upper limits. See Figure~\ref{fig:size_hue_bin} for binned data.}
    \label{fig:size_hue}
\end{figure*}

We simultaneously model the rms and phase-lag spectra of the QPO, and the time-averaged energy spectrum of the source for each observation to measure the corona size. Figure~\ref{fig:size_hue} presents the dependence of the corona size, modeled by \vk, upon hue, illustrating how the spatial extent of the corona changes across different spectral states of BHXRBs. 
The figure is divided into four spectral states, as previously shown in Figures~\ref{fig:pcd} and~\ref{fig:qpo_hue}.

\subsubsection{Trends and source-specific observations}

In the LHS~\citep[$\text{hue} < 140^\circ$][]{2015MNRAS.448.3339H}, which is dominated by emission from inverse Compton scattering in the corona~\citep{2006ARA&A..44...49R}, sources exhibit the largest corona sizes, typically ranging from $\sim$$10^4$ to $10^5$ km. The data points for GRS~1915+105 cluster around the hue of $50^\circ$--$80^\circ$, with the corona size increasing from $6\times10^3$ km to a few times $10^4$ km. As the hue increases from $50^\circ$ to $140^\circ$, data points from GX~339$-$4, MAXI~J1820+070, and Swift~J1727.8$-$1613 indicate that the corona size continues to grow, reaching up to $10^5$ km. Only one data point from \exo\ is located in the LHS at a hue of $\sim 110^\circ$, with the corona size constrained to be less than $\sim 10^5$ km.

As sources transition into the HIMS ($140^\circ < \text{hue} < 220^\circ$), the corona size decreases significantly, typically falling below $10^4$ km. In fact, the contraction of the corona already begins in the LHS when the hue approaches $140^\circ$. This decrease is particularly evident in \exo, MAXI~J1820+070, and Swift~J1727.8$-$1613. In contrast, for MAXI~J1535$-$571 and MAXI~J1631$-$479, the corona size ranges between 200 km and $10^4$ km within a narrow hue range. Near the transition from the HIMS to the SIMS, the corona size increases again, from several hundred kilometers (MAXI~J1803$-$298, MAXI~J1348$-$630) to a few thousand kilometers (GRS~1915+105, MAXI~J1348$-$630, and MAXI~J1820+070). 
Note that near the HIMS-to-SIMS transition, the data points are much sparser due to the rapid nature of the transition in BHXRBs, resulting in fewer observational snapshots.
The SIMS ($220^\circ < \text{hue} < 300^\circ$) is characterized by relatively compact coronae, with sizes typically below $10^4$ km. 
In the SIMS, the dataset consists of a few observations,
where QPO is less frequently observed, consistent with findings for the HIMS in other studies~\citep[e.g.,][]{2005A&A...440..207B,2018ApJ...865L..15S,2023MNRAS.520.5144Z,2023MNRAS.525..854M}.
In the HSS ($\text{hue} > 300^\circ$), there is only one observation from MAXI~J1348$-$630 with corona size of $\sim$8000~km. Note that the error bar of the hue marks this observation consistent with the SIMS. 

\begin{figure}
    \includegraphics[width=0.45\textwidth]{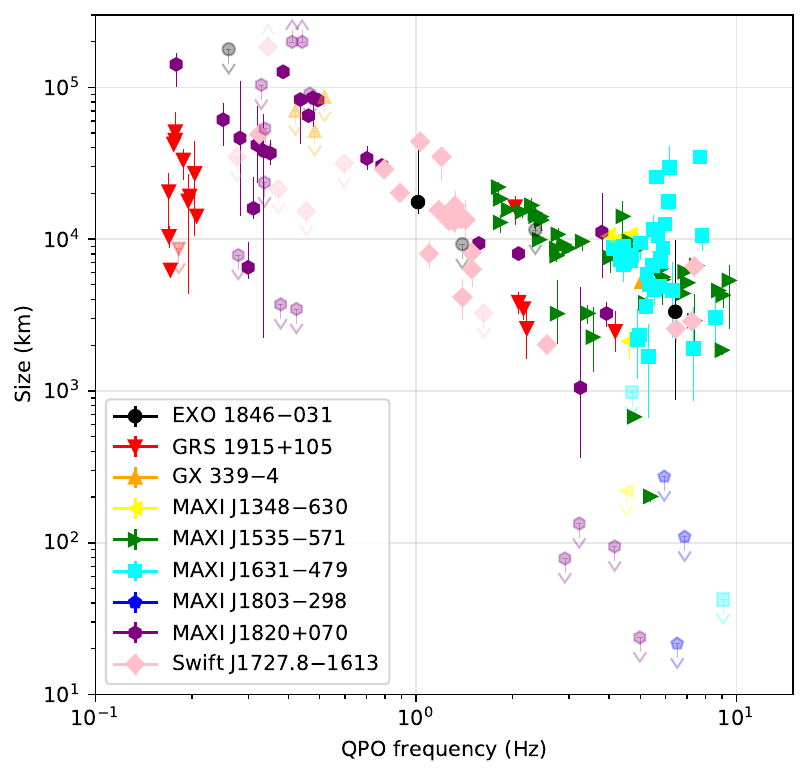}
    \caption{Coronal size as a function of the QPO frequency for the nine sources. The error bars indicate 1-$\sigma$ uncertainties. The points with arrows indicate the 90\% lower or upper limits.}
    \label{fig:size_qpo}
\end{figure}

Figure~\ref{fig:size_qpo} shows the corona size as a function of the QPO frequency. The coronal size first increases from $\sim10^4$~km up to $10^5$~km as the QPO frequency increases from $\sim$0.2~Hz to 0.5~Hz, and then decreases below $10^3$~km with increasing QPO frequency over the range 0.5--10~Hz. The decrease in coronal size exhibits a substantial range, similar to that observed within a narrow range of power-color hue in Figure~\ref{fig:size_hue}.

\subsubsection{Averaged corona size evolution and feedback fraction}

\begin{figure*}
    \includegraphics[height=0.45\textwidth]{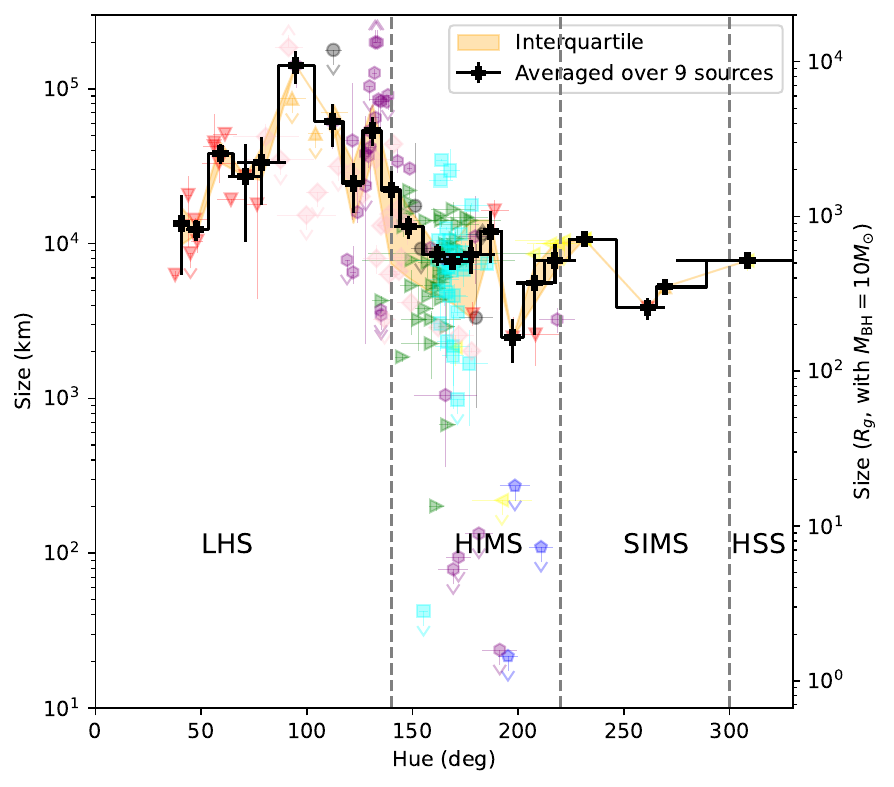}
    \includegraphics[height=0.45\textwidth]{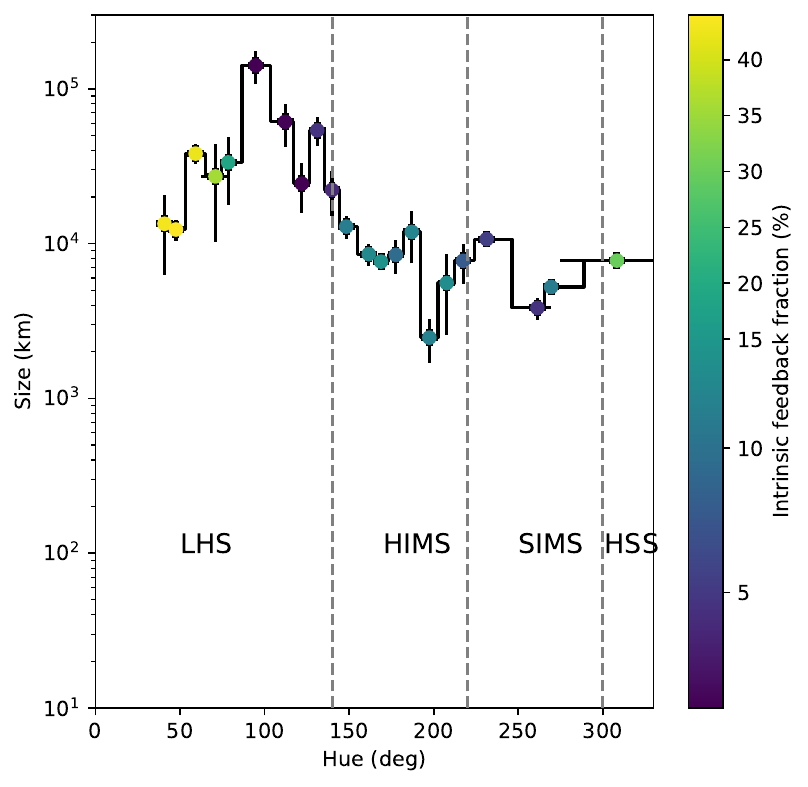}
    \caption{Averaged corona size as a function of hue for nine BHXRBs. Left: Individual measurements for each sources are given by the colored points (see legends in previous Figure), while the black curve represents the averaged corona size at each hue. The shaded orange region indicates the interquartile range (25\%–-75\%) of the corona size distribution within each bin. Right: The same averaged trend, with colors indicating the intrinsic feedback fraction (the fraction of corona photons that return to the accretion disk). In both panels, the vertical dashed lines separate the LHS, HIMS, SIMS, and HSS.}
    \label{fig:size_hue_bin}
\end{figure*}

To better resolve the evolution of corona size as a function of hue, we bin the data into intervals of $10^\circ$ in hue. Figure~\ref{fig:size_hue_bin} presents the evolution of corona size after averaging over the nine BHXRBs. The left panel shows individual data points from multiple sources, along with a black curve representing the averaged trend of corona size as a function of hue. Error bars of the black points denote the propagated uncertainty of the mean within each bin. The right panel displays the same averaged corona size, color-coded by the intrinsic feedback fraction, $\eta_{\text{int}}$, indicating the fraction of the corona photons that return to the accretion disk and are thermally reprocessed there.

The general trend reveals fluctuations in corona size as sources transition from the LHS to the HIMS and further into the SIMS. In the LHS ($\text{hue} < 140^\circ$), the corona remains large, with sizes increasing by about an order of magnitude from $\sim$$10^4$ km to $\sim$$10^5$ km at a hue of $\sim$$100^\circ$. As the hue increases toward the HIMS ($140^\circ < \text{hue} < 220^\circ$), the corona contracts significantly by about two orders of magnitude, with typical sizes decreasing to $\sim$$10^3$ km. This shrinkage is followed by an expansion shortly before the transition to the SIMS ($220^\circ < \text{hue} < 300^\circ$), where the corona reaches its most compact form. Only a few observations exist in the SIMS and marginally HSS, but the data suggest a relatively stable, small corona, with corona size variations between 4000 and 8000~km rather than the broader range of $\sim$$10^3$--$10^5$~km observed in harder states.

Further exploration of the intrinsic feedback fraction provides additional insights by linking corona size to the fraction of Comptonized photons reprocessed in the accretion disk (right panel of Figure~\ref{fig:size_hue_bin}). Darker colors indicate lower intrinsic feedback fractions, while lighter colors correspond to higher fractions. In the LHS, the intrinsic feedback fraction decreases from $\sim$40\% to $<1\%$ as the corona size increases, implying that most Comptonized photons escape without significant interaction with the disk when the corona size is the largest. However, as the system evolves from the LHS into the HIMS, the corona contracts dramatically while the intrinsic feedback fraction increases ($\sim$$10$--$15\%$), suggesting stronger coupling between the corona and the accretion disk. During the HIMS-to-SIMS transition, the corona size increases again by approximately a factor of three, while the intrinsic feedback fraction simultaneously drops to $\lesssim 5\%$. This behavior indicates a weakened interplay between the disk and corona, which is commonly observed during the HIMS-to-SIMS transition~\citep{2022ApJ...930...18W,2023MNRAS.525..854M}. 
As the source further evolves into the margin of SIMS/HSS, the feedback fraction increases to $\sim$30\%, suggesting an enhanced coupling between the corona and the disk.

\section{Discussion}\label{sec:discussion}

We have analyzed all \nicer\ archival data of nine bright BHXRBs with robust measurements of QPOs. A total of 171 QPO observations were identified across different spectral states, classified on the basis of the hue values in a power-color diagram (PCD). The QPO frequency increases from $\sim$0.1 to 10~Hz in the LHS and HIMS and generally remains constant at $\sim$4--5~Hz in the SIMS, while the hue continues to increase. Using the \texttt{vKompth} model~\citep{2020MNRAS.492.1399K,2022MNRAS.515.2099B}, we characterized the evolution of the corona geometry across states and QPO frequencies (see Figures~\ref{fig:size_hue}, \ref{fig:size_qpo} and~\ref{fig:size_hue_bin}). However, not all sources contribute data in all spectral states, and in some hue ranges the average is dominated by a single sampled source. The averaged profiles should therefore be interpreted as indicative population trends rather than as unbiased ensemble properties.

\subsection{Corona evolution during state transition}

The scatter in the corona size at similar values of hue is further resolved by the size versus QPO frequency. Actually both hue and QPO frequency provide useful tracers of the accretion geometry, when the source undergoes rapid and complex state transitions~\citep{2015MNRAS.448.3339H,2016ASSL..440...61B}. Changes in the corona size and coupling to the disk may occur on time-scales of days to weeks~\citep[e.g.][]{2023MNRAS.520..113R,2023MNRAS.525..854M}, leading to significant variations even within a narrow range of hue or QPO frequency. In addition, neither the hue nor the QPO frequency determines the corona geometry. The observed scatter thus likely reflects evolution of the corona rather than measurement uncertainties.

For a look at the corona size across state transitions, the corona size first increases and then decreases as sources transition from the LHS to the HIMS, and then into the SIMS. In the LHS ($\text{hue} < 140^\circ$), the corona is large, growing by about an order of magnitude from $\sim$$10^4$~km to above $10^5$~km around a hue of $\sim$$100^\circ$. As the source evolves toward the HIMS ($140^\circ < \text{hue} < 220^\circ$), the corona undergoes a rapid contraction, with sizes decreasing by roughly two orders of magnitude, down to $\sim$$2\times10^3$~km at the hue of $\sim200^{\circ}$. This is followed by a flare-like expansion near the HIMS-to-SIMS transition ($220^\circ < \text{hue} < 300^\circ$). In the SIMS, the corona becomes compact and relatively stable, with sizes ranging between 4000 and 8000~km. Although the SIMS and HSS are less sampled, the data indicate a narrower distribution in coronal size compared to the broader range of $\sim$$10^3$--$10^5$~km observed in harder states. We note that these evolutionary trends are traced as continuous functions of hue; the labels ``LHS'', ``HIMS'', and so on are used only as guides and do not affect the results, which are independent of the exact choice of state boundaries or classification method (e.g.\ power color versus hardness–intensity diagram).

Additionally, the intrinsic feedback fraction remains relatively low during the corona expansion in the LHS, possibly corresponding to ejection behavior of a large corona, as we have already found in other sources with \vk~\citep[e.g.][]{2022NatAs...6..577M,2023MNRAS.520.5144Z,2023MNRAS.525..854M}. These observed trends can be closely connected to spectral state transitions, where the contribution from the corona to the total emission in the $\sim$0.2--1000~keV decreases while that from the accretion disk increases~\citep{2007A&ARv..15....1D}.

As shown in our measurements, the corona size increases as the source evolves in the LHS, going from $\sim$$10^4$~km to nearly $10^5$~km as the hue increases and the source evolves toward the HIMS, while the intrinsic feedback fraction decreases. This behavior may be associated with the presence of a steady compact jet, which is commonly observed in the LHS~\citep{2006csxs.book..381F}. The expanding corona may play a role in supplying energy or material to the jet, or it may be partially extended along the jet axis, and be coupled weakly with the disk, resulting in fewer photons returning to the disk~\citep{2021NatCo..12.1025Y,2021A&A...646A.112R,2023Sci...381..961Y}. Additionally, the accretion disk in the LHS is relatively cool and likely truncated far from the black hole~\citep{2004MNRAS.351..791Z,2008NewAR..51..733N}, reducing the solid angle subtended by the disk from the corona's point of view. As a result, a significant fraction of Comptonized photons may escape without being intercepted and reprocessed by the disk, contributing to the observed decrease in the intrinsic feedback fraction. This scenario suggests a weak coupling between the corona and the disk in the LHS.

The flare-like expansion of the corona and the drop in corona feedback onto the disk seen near the HIMS-to-SIMS transition is an important feature. One possible explanation is that this marks the beginning of a short-lived ejection event, where part of the corona is expelled to form or feed a relativistic jet~\citep{2022NatAs...6..577M}. This idea fits with observations showing that steady jets disappear and fast-moving ejecta are launched during this phase~\citep{1994Natur.371...46M,2009MNRAS.396.1370F}. In this case, the sudden expansion of the corona may reflect changes in magnetic fields or how the corona interacts with the disk, making it easier for material to be ejected upward~\citep{2006A&A...447..813F}. 
In contrast, during the LHS-to-HIMS transition, we observe a steady contraction of the corona accompanied by an increase in the intrinsic feedback fraction. This trend supports the scenario in which the corona collapses while increasing its interaction with the disk~\citep{2019Natur.565..198K,2022NatAs...6..577M}, leading to enhanced disk reprocessing in relatively softer states.

Recent X-ray polarization measurements with \textit{IXPE} provide additional constraints on the corona geometry in BHXRBs, favoring a corona structure that is predominantly projected in the disk plane during the LHS and HIMS, with only modest changes in polarization angle~\citep{2022Sci...378..650K,2024ApJ...968...76I}. These results are complementary to our timing-based study, which probes the corona size, evolution, and coupling with the disk and/or jet, but do not determine its orientation. In this context, a large coronal structure along the disk plane, potentially associated with the jet base, can remain consistent with polarization. The rapid contraction of the corona size observed as the sources evolve from the LHS toward the HIMS occurs contemporaneously with the decrease of the polarization degrees observed in some of these sources, suggesting a varying corona geometry~\citep{2023ApJ...958L..16V,2024ApJ...968...76I}.

\subsection{Origin of the seed photons}\label{discussion:seed photons}

The evolution of the ratio between the disk temperature, $kT_{\rm in}$, in \texttt{diskbb} and the seed photon temperature, $kT_{\rm s}$, in \vk\ provides insight into the changing geometry and coupling between the accretion disk and the corona. Since this ratio directly compares the characteristic temperature of the inner disk with that of the seed photons within the \vk\ framework, deviations
from unity indicate that the seed photons do not originate from the same disk region described by the \texttt{diskbb} component.

In the LHS, the ratio $kT_{\rm in}/kT_{\rm s}$ significantly deviates from unity, suggesting that the seed photons in the variable Comptonization are not directly from the inner disk region dominating the time-averaged thermal emission. Instead, the seed photons may originate from cooler regions of the disk~\citep[e.g.][]{2013MNRAS.430.3196V}. In the meanwhile, the intrinsic feedback fraction drops significantly from $\sim$40\% to $<1\%$. This interpretation is consistent with a relatively extended corona whose coupling to the inner disk gradually weakens in the hard state.
As the source evolves toward the HIMS, the ratio gradually approaches unity while the intrinsic feedback fraction increases, indicating an increasing coupling between the corona and the inner disk. This behavior is simultaneous with the contraction of the corona size and with the increase in the intrinsic feedback fraction, consistent with a more compact disk-corona geometry~\citep[e.g.][]{1997ApJ...489..865E}.
During the transition from the HIMS to the SIMS, the ratio increases and the intrinsic feedback fraction decreases again, implying a decoupling between the inner disk and the seed photon source. This behavior may reflect a change in the coronal geometry or heating mechanism during the rapid state transition, known to be accompanied by substantial changes in accretion flow properties~\citep{2004MNRAS.355.1105F,2005A&A...440..207B}
From the SIMS to the HSS, the ratio decreases and evolves towards unity, indicating that the seed photons are again closely associated with the inner disk in the soft state. In this regime, the system appears to reach a more stable configuration, where the thermal disk dominates the emission and much less observations show QPOs~\citep[e.g.][]{2011MNRAS.418.2292M,2024MNRAS.527.5638Z}.

\subsection{Corona size, black hole mass and disk inclination}

While the evolution of corona size is robustly measured, it is important to consider whether such trends could be biased by black hole mass or inclination of the system, or the assumptions in the model.

Many black hole properties scale with the black hole mass, for instance the radius of the ISCO, or the Eddington luminosity~\citep{2010ApJ...718L.117S,2014ApJ...782...76G,2022NatAs...6.1364G}. It is therefore crucial to consider the black hole masses in our sample. Only five sources among the nine BHXRBs analyzed here have dynamical mass measurements: GRS~1915+105 ($11.2 \pm 2~M_{\odot}$; \citealt{2023ApJ...959...85R}), GX~339$-$4 (ranging from 2.3 to 9.5~$M_{\odot}$; \citealt{2017ApJ...846..132H}), MAXI~J1803$-$298 (3 to 10~$M_{\odot}$; \citealt{2022ApJ...926L..10M}), MAXI~J1820+070 (5.7 to 9.2~$M_{\odot}$; \citealt{2020ApJ...893L..37T}), and Swift~J1727.8$-$1613 ($> 4~M_{\odot}$; \citealt{2025A&A...693A.129M}). These values are broadly consistent with each other and with the typical $\sim$9--10~$M_{\odot}$ expected for stellar-mass black holes. For the remaining sources, direct dynamical estimates are not yet available, and we adopt a representative value of $10~M_{\odot}$ for scaling purposes. Therefore, we choose to present the corona size in physical units (kilometers), as directly obtained from \vk, and scale it into $R_g$ assuming a fiducial mass of $10~M_{\odot}$ (see the right axis of the left panel in Figure~\ref{fig:size_hue_bin}).

\citet{2017MNRAS.464.2643V} reported that the sign (positive or negative) of the QPO phase lag, when the QPO frequencies are above $\sim$2~Hz, depends on the inclination of the BHXRB, possibly due to differences in relativistic geometrical effects such as Lense–Thirring precession. In our \vk\ framework, QPOs are modeled as oscillations in thermal properties, i.e.\ temperature, in the corona. The model accounts for photon propagation through both direct and inverse Comptonization in the corona, as well as thermal reprocessing in the disk, all of which contribute to the observed changes in phase lags. The sources in our sample span a wide range of inclination angles: EXO~1846$-$031 ($\sim$$40^\circ$; \citealt{2021ApJ...906...11W}), GRS~1915+105 ($\sim$$70^\circ$; \citealt{2013ApJ...775L..45M}), GX~339$-$4 ($48^\circ \pm 1^\circ$; \citealt{2015ApJ...813...84G}), MAXI~J1348$-$630 ($<46^\circ$; \citealt{2021MNRAS.504..444C}), MAXI~J1535$-$571 ($\sim$$60^\circ$; \citealt{2018ApJ...852L..34X}), MAXI~J1631$-$479 ($29^\circ \pm 1^\circ$; \citealt{2020ApJ...893...30X}; See~\citealt{2023ApJ...944...68R} for other measurements of $50^\circ$--$70^\circ$), MAXI~J1803$-$298 ($\sim$$70^\circ$; \citealt{2022MNRAS.516.2074F}), MAXI~J1820+070 ($81^\circ$; \citealt{2021ApJ...910L...3W}), and Swift~J1727.8$-$1613 ($\sim$$40^\circ$; \citealt{2024ApJ...960L..17P}). Despite the wide variation in inclination, our results suggest that the inferred corona size evolution does not strongly depend on viewing angle. For instance, all sources exhibit a broadly similar pattern: an initial expansion in the LHS, a significant contraction in the HIMS, and a flare-like expansion near the HIMS-to-SIMS transition. The SIMS is characterized by a more compact and stable corona with reduced variability, while around the HIMS to HSS transition, though sparsely sampled due to few QPO detections, shows tentative signs of a slight increase in corona size. We note, however, that not all sources are equally represented across all states, so this description reflects the global picture rather than a uniform sequence observed in every case. This inclination-independent trend is consistent with the assumptions of \texttt{vKompthdk}, which models the corona as spherical and isotropic, and which likely indicates that inclination does not have a significant effect on the inferred corona size.

\subsection{Coronal geometry from different timing signals}

Based on fits using the \vk\ model, we have previously investigated \nicer\ observations of the corona evolution during state transitions in several BHXRBs, including MAXI~J1348$-$630~\citep{2021MNRAS.501.3173G,2025ApJ...980..251A}, MAXI~J1535$-$571~\citep{2023MNRAS.520..113R,2023MNRAS.526.3944Z}, MAXI~J1820+070~\citep{2023MNRAS.525..854M}, and Swift~J1727.8$-$1613~\citep{2025A&A...697A.229R}. For instance, the study of MAXI~J1820+070 focuses on the variable corona during the HIMS-to-SIMS transition over timescales of hours, showing a morphological change from a slab-like to a jet-like corona geometry~\citep{2023MNRAS.525..854M}. In MAXI~J1535$-$571, QPOs are detected in the HIMS at frequencies between 2 and 9~Hz, where the corona size continuously decreases from $\sim$$10^4$ to $\sim$$10^3$~km and the feedback fraction increases~\citep{2023MNRAS.520..113R}. Note that this behavior is observed within a narrow hue range of $\sim$$145^\circ$--$175^\circ$, corresponding only to the HIMS. These earlier works concentrate on specific spectral states, rather than a systematic view of corona evolution across full outbursts or larger source samples. In contrast, our study aims to provide such an overview by examining the complete outburst evolution of nine BHXRBs, with corona measurements in individual \nicer\ orbits across all spectral states.

A recent study of Swift~J1727.8$-$1613 using a dual-corona model finds that the overall corona size is consistent with that inferred from the one-zone \vk\ model~\citep{2025A&A...697A.229R}. In that work, the corona size increases to $\sim$$10^4$~km in the LHS and subsequently decreases to $\sim$$2\times 10^3$~km in the HIMS. Meanwhile, the intrinsic feedback fraction shows a clear increase during the LHS-to-HIMS transition, consistent with our findings for Swift~J1727.8$-$1613.

While X-ray variability in BHXRBs generally comprises broadband noise and QPOs~\citep{2002ApJ...572..392B}, recent works suggest a different interpretation of the PDS. Several studies~\citep{2000MNRAS.318..361N,2002ApJ...572..392B,2024MNRAS.527.9405M} propose that the PDS is better described as a sum of multiple Lorentzian components, each representing variability on relatively narrow timescales. At low frequencies, broadband noise often exhibits hard lags attributed to accretion fluctuations and Comptonization~\citep{2025MNRAS.536.3284U}, while at higher frequencies, soft lags interpreted as soft delays may arise from disk reflection~\citep{2010MNRAS.401.2419Z}. Models like \texttt{reltrans} assume a lamppost geometry to describe such reverberation, but treat reverberation and hard lags as being produced by distinct mechanisms~\citep{2021ApJ...910L...3W}. In our study, we focus on constraining coronal properties using the QPO component, which represents a coherent feature in the PDS and cross spectrum~\citep{2024MNRAS.527.9405M}. The lags of QPOs can be either hard due to Comptonization or soft due to feedback thermalization~\citep{2020MNRAS.492.1399K,2022MNRAS.515.2099B}. By modeling the QPO phase-lag and rms spectra of the QPO jointly with the energy spectrum of the source, our approach isolates a specific mode of coronal variability, allowing us to directly probe the physical structure and evolution of the corona across spectral states.

Interestingly, the soft lags inferred from broadband noise studies change significantly during spectral state transitions~\citep{2022ApJ...930...18W}. Assuming these soft lags arise from the light travel time between the corona and the disk, estimates suggest that during the hard-to-intermediate state transition, the corona height decreases from $\sim$1000 km to 200 km in the LHS, then expands up to $\sim$9000 km in the HIMS, followed by a slight decrease in the SIMS. This trend is similar to the behavior we observe from QPO modeling, where corona size contracts in the HIMS and flares near the HIMS-to-SIMS transition. However, the absolute size estimates differ: our \texttt{vKompthdk} modeling shows the corona reaching up to $10^5$~km in the LHS, dropping to $\sim$$10^3$~km in the HIMS, and remaining compact at $\sim$4000--8000~km in the SIMS.


An important distinction should be emphasized between the corona sizes inferred from QPO modeling and those estimated from reverberation soft lags, which are commonly interpreted within a lamppost geometry. Reverberation lags primarily trace light-travel time between a vertical point X-ray source and the disk~\citep{2014A&ARv..22...72U}, whereas \vk\ constrains an effective corona size associated with Comptonization and disk thermalization. In the early LHS, the large feedback fraction (40\%) and corona size inferred from QPO modeling are naturally explained by an extended corona strongly coupled to the disk, rather than vertically elevated lamppost~\citep{2025ApJ...980..251A}. In such a configuration, large feedback in \vk\ does not necessarily produce large reverberation soft lags, highlighting the different physical sensitivities of the two timing diagnostics.

Within this framework, the two approaches should be regarded as complementary, each probing different aspects of the corona geometry under different assumptions.
Despite differences in scale, both reverberation and QPO-based analyses reveal similar trends in coronal geometry. Prior to the HIMS-to-SIMS transition, the soft lags of the broadband noise and the corona size inferred from the QPO using \texttt{vKompthdk} modeling both suggest that the corona becomes more compact, or that disk-corona coupling strengthens~\citep{2022ApJ...930...18W}. However, around the HIMS-to-SIMS transition, the two diagnostics appear to decouple: the corona height inferred from soft lags increases significantly, while the \texttt{vKompthdk}-inferred size remains compact. This divergence likely reflects a geometric change, such as the vertical expansion or partial ejection of the corona. In such a scenario, in the LHS the highly coherent QPO signal may catch the full size of an extended Comptonization medium, while the reverberation lag would be sensitive to an ``averaged'' coronal height, or base of the vertical corona, from where most of the photons are emitted towards the disk producing the soft lag. This interpretation is supported by simulations~\citep{2021MNRAS.507..983L,2023ApJ...951...19L} and reflection-based modeling~\citep{2019MNRAS.490.1350B}, which suggest that the corona transitions from a vertically extended structure in the LHS to a more compact, possibly jet-associated configuration in a softer state. The discrepancy in absolute size estimates may thus be reconciled by recognizing that reverberation lags are influenced not only by light travel time, but also by scattering, dilution, and energy-dependent response. A unified modeling approach incorporating both QPO and reverberation timing may offer a more complete picture of the evolving coronal geometry throughout the outburst.




\section*{Acknowledgements}

We thank the referee for their constructive comments that help to improve the quality of this paper.
We thank David M.\ Russell for helpful discussions.
YZ acknowledges support from the Dutch Research Council (NWO) Rubicon Fellowship, file no.\ 019.231EN.021.
MM acknowledges support from the research programme Athena with project number 184.034.002, which is (partly) financed by NWO.
FG is a CONICET researcher and acknowledges support by PIP 0113 and PIBAA 1275 (CONICET). FG acknowledges support from PIBAA 1275 (CONICET). FG was also supported by grant PID2022-136828NB-C42 funded by the Spanish MCIN/AEI/ 10.13039/501100011033 and ``ERDF A way of making Europe''.
KA and SKR are supported by Tamkeen under the NYU Abu Dhabi Research Institute grant CASS.


\section*{Data Availability}

The X-ray data used in this article are accessible at NASA's High Energy Astrophysics Science Archive Research Center~\url{https://heasarc.gsfc.nasa.gov/}. The software GHATS for Fourier timing analysis is available at~\url{http://www.brera.inaf.it/utenti/belloni/GHATS_Package/Home.html}.



\bibliographystyle{mnras}
\bibliography{nicer_reference} 



\appendix

\section{Observations}\label{sec: appendix observations}

\clearpage
\onecolumn
\begin{longtable}{clcccccc}
\caption{The \nicer\ observations used for the joint fitting. In the orbital column, ``F'' denotes the full observation, while numbers indicate the \nicer\ orbits that we use. The corona size ranges from $0.01 \times 10^3$~km to $200 \times 10^3$~km, while the feedback fraction ranges from 0 to 1, as set by the model. The quoted uncertainties correspond to $1\sigma$. Note that the symbol ``p'' in the error bars indicates that the parameter is pegged at the boundary on that side. In such cases, the opposite side represents the 90\% confidence interval.}\label{tab:obs}\\
\hline
Name & \nicer\ ID & Orbit & QPO frequency (Hz) & Hue (deg) & Size ($10^3$~km) & Feedback fraction & $\chi^2_{\text{reduced}}$ \\
\hline
EXO~1846$-$031 & 2200760101 & F & $0.26\pm 0.01$ & $112\pm 3$ & $0.01^{+178}_{-p}$ & $1.0^{+p}_{-0.5}$ & 0.52\\
 & 2200760103 & F & $1.01\pm 0.02$ & $151\pm 3$ & $18^{+27}_{-3}$ & $0^{+0.9}_{-p}$ & 0.47\\
 & 2200760104 & F & $1.39\pm 0.02$ & $153\pm 4$ & $2^{+7}_{-p}$ & $0.0^{+0.5}_{-p}$ & 0.61\\
 & 2200760105 & F & $2.36\pm 0.02$ & $183\pm 7$ & $8^{+p}_{-p}$ & $0.0^{+0.5}_{-p}$ & 0.48\\
 & 2200760110 & F & $6.4\pm 0.2$ & $179\pm 8$ & $3^{+7}_{-2}$ & $0.5^{+p}_{-0.2}$ & 0.31\\
\hline
GRS~1915+105 & 1103010137 & F & $4.18\pm0.07$ & $197\pm 4$ & $2.5^{+0.9}_{-0.7}$ & $0.48\pm 0.06$ & 0.68\\
 & 1103010157 & F & $2.16\pm 0.01$ & $178\pm 1$ & $3.5\pm0.7$ & $0.41\pm +0.07$ & 0.60\\
 & 2596010201 & F & $2.04\pm 0.01$ & $188\pm 6$ & $16\pm4$ & $0.8^{+p}_{-0.3}$ & 1.04\\
 & 2596010601 & F & $2.21\pm 0.01$ & $208\pm 11$ & $3\pm1$ & $0.32\pm 0.02$ & 1.39\\
 & 2596010801$^{**}$ & F & $2.09\pm0.02$ & $261\pm 8$ & $3.8\pm 0.6$ & $0.70\pm 0.03$ & 1.60\\
 & 4103010225$^{*,\ **}$ & F & $0.170\pm 0.002$ & $48\pm 2$ & $10\pm 2$ & $0.97\pm 0.01$ & 1.44\\
 & 4647012001$^{*,\ **}$ & F & $0.182\pm 0.002$ & $44\pm 1$ & $0.01^{+9}_{-p}$ & $0.94\pm 0.01$ & 1.37\\
 & 4103010226$^{*,\ **}$ & F & $0.175\pm0.004$ & $56\pm 5$ & $42\pm3$ & $1.0^{+p}_{-0.1}$ & 1.27\\
 & 4647012002$^{*,\ **}$ & F & $0.171\pm 0.002$ & $37\pm 1$ & $6.3\pm0.4$ & $0.2^{+0.2}_{-p}$ & 1.49\\
 & 4103010227$^{*,\ **}$ & F & $0.169\pm 0.002$ & $43\pm 1$ & $21^{+7}_{-11}$ & $1.00^{+p}_{-0.06}$ & 1.62\\
 & 4103010230$^{*,\ **}$ & F & $0.177\pm0.003$ & $61\pm 5$ & $51.0^{+0.6}_{-0.9}$ & $0.92^{+0.02}_{-0.19}$ & 1.43\\
 & 4103010231$^{**}$ & F & $0.177\pm 0.002$ & $56\pm 3$ & $45^{+24}_{-3}$ & $0.93^{+p}_{-0.09}$ & 1.19\\
 & 4103010235$^{**}$ & F & $0.195\pm 0.002$ & $76\pm 4$ & $18^{+9}_{-14}$ & $0.86^{+0.06}_{-0.18}$ & 1.61\\
 & 4103010236$^{**}$ & F & $0.188\pm 0.002$ & $58\pm 7$ & $33^{+6}_{-9}$ & $0.96^{+p}_{-0.05}$ & 1.53\\
 & 4103010237$^{*,\ **}$ & F & $0.196\pm0.004$ & $64\pm 6$ & $19.3^{+0.1}_{-1.0}$ & $0.96^{+p}_{-0.05}$ & 1.44\\
 & 4103010238$^{*,\ **}$ & F & $0.207\pm 0.002$ & $46\pm 5$ & $14.3^{+0.4}_{-0.2}$ & $0.96^{+p}_{-0.05}$ & 1.58\\
 & 4647012201$^{**}$ & F & $0.204\pm 0.003$ & $70\pm 8$ & $27\pm 17$ & $0.96^{+p}_{-0.05}$ & 1.32\\
\hline
GX~339$-$4 & 3558011402 & F & $0.42\pm 0.01$ & $113\pm 6$ & $0.01^{+70}_{-p}$ & $1.0^{+p}_{-0.6}$ & 0.83\\
 & 3558011501 & F & $0.48\pm 0.01$ & $104\pm 3$ & $0.01^{+51}_{-p}$ & $0.8^{+p}_{-0.3}$ & 0.76\\
 & 4133010101 & F & $0.52\pm0.02$ & $92\pm 6$ & $0.01^{+86}_{-p}$ & $0.5^{+p}_{-0.2}$ & 0.70\\
 & 4133010107 & F & $5.05\pm 0.02$ & $269\pm 4$ & $5.2\pm0.5$ & $0.33\pm0.05$ & 1.63\\
\hline
MAXI~J1348$-$630 & 1200530107 & F & $4.62\pm 0.02$ & $170\pm 2$ & $2.1\pm0.5$ & $0.38\pm0.02$ & 1.46\\
 & 1200530108$^{*}$ & F & $4.53\pm 0.01$ & $192\pm 14$ & $0.01^{+0.22}_{-0.01}$ & $0.34\pm 0.01$ & 1.38\\
 & 1200530109$^{*}$ & F & $4.67\pm 0.01$ & $231\pm 5$ & $10.7^{+0.4}_{-0.1}$ & $0.85\pm 0.06$ & 1.68\\
 & 1200530112$^{*}$ & F & $4.51\pm 0.01$ & $218\pm 10$ & $10.0^{+0.5}_{-2.0}$ & $0.8\pm 0.1$ & 1.65\\
 & 1200530113 & F & $4.38\pm 0.03$ & $308\pm 33$ & $7.8^{+0.6}_{-1.1}$ & $1.0^{+p}_{-0.1}$ & 1.08\\
 & 1200530117$^{*}$ & F & $4.28\pm0.01$ & $215\pm 10$ & $10.0^{+0.4}_{-2.5}$ & $0.8^{+p}_{-0.1}$ & 1.58\\
 & 1200530123 & F & $3.97\pm0.04$ & $221\pm 23$ & $2^{+8}_{-p}$ & $0.2^{+0.2}_{-0.1}$ & 1.00\\
 & 1200530124$^{*}$ & F & $4.24\pm0.01$ & $207\pm 2$ & $9\pm1$ & $0.7\pm0.1$ & 1.29\\
 & 1200530125$^{*}$ & F & $3.98\pm0.01$ & $231\pm 9$ & $10.7\pm0.8$ & $0.00^{+0.02}_{-p}$ & 1.67\\
\hline
MAXI~J1535$-$571 & 1050360105 & 2 & $2.75\pm0.02$ & $152\pm 4$ & $3\pm2$ & $0.55\pm0.07$ & 0.70\\
 & 1050360105 & 3 & $2.42\pm0.01$ & $165\pm 3$ & $9.9\pm0.2$ & $1.00^{+p}_{-0.03}$ & 0.59\\
 & 1050360105 & 4 & $2.29\pm0.01$ & $169\pm 10$ & $17\pm3$ & $1.0^{+p}_{-0.2}$ & 0.81\\
 & 1050360105 & 5 & $2.35\pm0.01$ & $175\pm 10$ & $15^{+1}_{-3}$ & $1.0^{+p}_{-0.3}$ & 0.70\\
 & 1050360105 & 6 & $2.46\pm0.02$ & $171\pm 10$ & $13.4^{+0.9}_{-2.1}$ & $1.0^{+p}_{-0.1}$ & 0.73\\
 & 1050360105 & 11 & $1.94\pm0.01$ & $147\pm 3$ & $15.6\pm0.2$ & $1.00^{+p}_{-0.03}$ & 0.71\\
 & 1050360105 & 12 & $1.80\pm0.01$ & $148\pm 2$ & $22\pm2$ & $1.00^{+p}_{-0.07}$ & 0.69\\
 & 1050360105 & 13 & $1.83\pm0.01$ & $145\pm 2$ & $13\pm2$ & $0.59\pm0.07$ & 0.64\\
 & 1050360106 & F & $1.83\pm0.01$ & $148\pm 0$ & $18.5\pm0.8$ & $0.85^{+p}_{-0.08}$ & 0.39\\
 & 1050360107 & F & $2.14\pm0.01$ & $163\pm 2$ & $15\pm1$ & $1.0^{+p}_{-0.2}$ & 0.30\\
 & 1050360108 & 0 & $2.36\pm0.02$ & $163\pm 6$ & $14.2^{+0.6}_{-1.4}$ & $1.0^{+p}_{-0.2}$ & 0.61\\
 & 1050360108 & 1 & $2.45\pm0.02$ & $156\pm 5$ & $14.1\pm0.9$ & $0.77\pm0.05$ & 0.78\\
 & 1050360108 & 2 & $2.74\pm0.01$ & $175\pm 23$ & $7.8\pm0.8$ & $0.67\pm0.04$ & 0.87\\
 & 1050360109 & 0 & $2.77\pm0.01$ & $170\pm 7$ & $11^{+1}_{-3}$ & $0.9^{+p}_{-0.3}$ & 0.74\\
 & 1050360109 & 1 & $3.30\pm0.02$ & $167\pm 6$ & $9.6^{+0.6}_{-1.3}$ & $1.0^{+p}_{-0.2}$ & 0.68\\
 & 1050360109 & 2 & $3.56\pm0.04$ & $158\pm 13$ & $2.3^{+1.3}_{-0.9}$ & $0.62^{+0.11}_{-0.09}$ & 0.84\\
 & 1050360109 & 5 & $3.43\pm0.02$ & $158\pm 13$ & $3.2\pm0.05$ & $0.66\pm0.04$ & 0.82\\
 & 1050360109 & 6 & $2.99\pm0.01$ & $173\pm 44$ & $8.7^{+0.4}_{-0.6}$ & $1.00^{+p}_{-0.05}$ & 0.67\\
 & 1050360109 & 7 & $2.74\pm0.01$ & $164\pm 4$ & $9\pm1$ & $0.78^{+0.08}_{-0.06}$ & 0.57\\
 & 1050360110 & 0 & $2.89\pm0.01$ & $166\pm 4$ & $9.2\pm0.4$ & $0.75\pm0.03$ & 0.66\\
 & 1050360110 & 1 & $4.03\pm0.02$ & $157\pm 4$ & $8\pm2$ & $0.8^{+p}_{-0.2}$ & 0.68\\
 & 1050360110 & 2 & $4.79\pm0.02$ & $166\pm 4$ & $0.68\pm0.03$ & $0.321^{+0.001}_{-0.007}$ & 0.75\\
 & 1050360111 & F & $9.07\pm0.05$ & $135\pm 4$ & $4.3^{+0.3}_{-0.8}$ & $1.0^{+p}_{-0.1}$ & 0.24\\
 & 1050360112 & 1 & $7.38\pm0.06$ & $166\pm 4$ & $3\pm1$ & $0.63^{+0.18}_{-0.09}$ & 0.60\\
 & 1050360113 & 0 & $7.50\pm0.05$ & $162\pm 2$ & $6.69\pm0.05$ & $0.796\pm0.005$ & 0.68\\
 & 1050360113 & 1 & $8.99\pm0.04$ & $145\pm 3$ & $1.86^{+0.07}_{-0.14}$ & $0.76\pm0.02$ & 0.69\\
 & 1050360113 & 2 & $8.77\pm0.03$ & $157\pm 2$ & $4.59\pm0.05$ & $0.687\pm0.005$ & 0.58\\
 & 1050360113 & 4 & $9.47\pm0.03$ & $149\pm 5$ & $5^{+1}_{-3}$ & $0.9^{+p}_{-0.4}$ & 0.74\\
 & 1130360103 & 1 & $6.88\pm0.04$ & $165\pm 2$ & $6.1^{+0.2}_{-0.4}$ & $1.00^{+p}_{-0.06}$ & 0.54\\
 & 1130360104 & F & $5.39\pm0.02$ & $161\pm 3$ & $0.20\pm0.02$ & $0.402^{+0.001}_{-0.010}$ & 1.14\\
 & 1130360105 & F & $5.80\pm0.01$ & $164\pm 1$ & $6.4\pm0.5$ & $0.83^{+0.07}_{-0.04}$ & 0.23\\
 & 1130360106 & F & $6.80\pm0.04$ & $163\pm 2$ & $4.39\pm0.09$ & $0.68\pm0.01$ & 0.27\\
 & 1130360107 & F & $4.61\pm0.01$ & $169\pm 3$ & $9.8^{+0.4}_{-0.8}$ & $1.00^{+p}_{-0.05}$ & 0.81\\
 & 1130360108 & 0 & $4.83\pm0.02$ & $162\pm 6$ & $8.5^{+0.8}_{-1.0}$ & $1.0^{+p}_{-0.2}$ & 0.62\\
 & 1130360108 & 1 & $5.87\pm0.03$ & $159\pm 4$ & $5.4^{+0.2}_{-1.7}$ & $1.00^{+p}_{-0.03}$ & 0.73\\
 & 1130360110 & 0 & $7.03\pm0.06$ & $161\pm 4$ & $5.2\pm0.2$ & $1.00^{+p}_{-0.06}$ & 0.63\\
 & 1130360110 & 1 & $5.88\pm0.03$ & $162\pm 4$ & $5.6^{+0.3}_{-0.5}$ & $0.77^{+0.10}_{-0.04}$ & 0.64\\
 & 1130360112 & 2 & $4.36\pm0.03$ & $149\pm 23$ & $7.9^{+0.5}_{-0.3}$ & $1.00^{+p}_{-0.09}$ & 0.77\\
 & 1130360113 & 0 & $4.41\pm0.03$ & $177\pm 14$ & $14\pm4$ & $1.0^{+p}_{-0.2}$ & 0.81\\
 & 1130360113 & 3 & $5.19\pm0.02$ & $155\pm 4$ & $3.81\pm0.09$ & $1.00^{+p}_{-0.02}$ & 0.71\\
\hline
MAXI~J1631$-$479 & 1200500110 & F & $5.21\pm0.02$ & $171\pm 2$ & $3.6^{+0.6}_{-0.2}$ & $0.6\pm0.1$ & 0.89\\
 & 1200500111 & F & $4.16\pm0.02$ & $167\pm 3$ & $8.7^{+0.4}_{-0.9}$ & $1.00^{+p}_{-0.09}$ & 0.36\\
 & 1200500112 & F & $7.69\pm0.08$ & $163\pm 2$ & $34.7\pm0.7$ & $1.0^{+p}_{-0.1}$ & 0.46\\
 & 1200500113 & 2 & $5.63\pm0.03$ & $163\pm 5$ & $25.7\pm0.2$ & $0.0^{+0.3}_{-p}$ & 0.82\\
 & 1200500113 & 3 & $5.55\pm0.03$ & $169\pm 5$ & $5\pm1$ & $0.55\pm0.06$ & 0.89\\
 & 1200500113 & 4 & $5.97\pm0.04$ & $165\pm 4$ & $12.6\pm0.3$ & $1.00^{+p}_{-0.01}$ & 0.85\\
 & 1200500113 & 5 & $4.90\pm0.03$ & $169\pm 8$ & $2\pm1$ & $0.37\pm0.05$ & 0.88\\
 & 1200500113 & 6 & $4.69\pm0.02$ & $164\pm 2$ & $7.2\pm0.2$ & $1.00^{+p}_{-0.03}$ & 0.70\\
 & 1200500114 & F & $5.79\pm-0.02$ & $166\pm 1$ & $7.1^{+0.6}_{-0.2}$ & $0.9^{+p}_{-0.2}$ & 0.26\\
 & 1200500115 & 0 & $5.74\pm0.03$ & $166\pm 3$ & $4.9\pm0.3$ & $0.69\pm0.06$ & 0.72\\
 & 1200500115 & 1 & $6.18\pm0.04$ & $167\pm 4$ & $30^{+11}_{-4}$ & $1.0^{+p}_{-0.1}$ & 0.73\\
 & 1200500116 & 0 & $8.6\pm0.1$ & $163\pm 4$ & $3\pm1$ & $0.56\pm0.05$ & 0.70\\
 & 1200500116 & 4 & $7.33\pm0.08$ & $169\pm 5$ & $2^{+2}_{-1}$ & $0.29^{+0.09}_{-0.06}$ & 0.92\\
 & 1200500116 & 8 & $7.8\pm0.1$ & $162\pm 4$ & $10.6^{+0.9}_{-2.2}$ & $1.0^{+p}_{-0.2}$ & 0.75\\
 & 1200500117 & F & $9.08\pm0.06$ & $155\pm 2$ & $0.01^{+0.04}_{-p}$ & $0.696^{+0.003}_{-0.014}$ & 0.28\\
 & 1200500118 & 5 & $6.30\pm0.06$ & $167\pm 9$ & $5\pm2$ & $0.6^{+0.4}_{-0.2}$ & 0.80\\
 & 1200500118 & 7 & $4.97\pm0.03$ & $166\pm 5$ & $2.3\pm0.2$ & $1.00^{+p}_{-0.02}$ & 0.69\\
 & 1200500118 & 8 & $5.32\pm0.03$ & $162\pm 5$ & $5^{+0.4}_{-0.6}$ & $1.00^{+p}_{-0.07}$ & 0.80\\
 & 1200500119 & 0 & $4.70\pm0.03$ & $161\pm 10$ & $7.8\pm0.06$ & $1.0^{+p}_{-0.1}$ & 0.95\\
 & 1200500119 & 2 & $4.34\pm0.03$ & $174\pm 10$ & $7.3\pm0.4$ & $1.00^{+p}_{-0.05}$ & 0.83\\
 & 1200500119 & 3 & $4.44\pm0.02$ & $173\pm 8$ & $6.8^{+0.9}_{-1.6}$ & $1.0^{+p}_{-0.3}$ & 0.78\\
 & 1200500119 & 4 & $4.46\pm0.02$ & $167\pm 6$ & $7.0\pm0.4$ & $1.00^{+p}_{-0.06}$ & 0.80\\
 & 1200500124 & 0 & $5.25\pm0.04$ & $165\pm 6$ & $6\pm2$ & $0.0^{+0.2}_{-p}$ & 0.89\\
 & 1200500126 & F & $4.13\pm0.03$ & $170\pm3$ & $9\pm2$ & $1.0^{+p}_{-0.3}$ & 0.27\\
 & 1200500127 & F & $4.43\pm0.01$ & $168\pm 2$ & $9^{+0.6}_{-0.9}$ & $1.0^{+p}_{-0.1}$ & 0.25\\
 & 1200500128 & F & $5.01\pm0.04$ & $164\pm 3$ & $9\pm2$ & $0.9^{+p}_{-0.4}$ & 0.33\\
 & 1200500129 & F & $5.90\pm0.05$ & $174\pm 5$ & $9\pm1$ & $1.0^{+p}_{-0.1}$ & 0.48\\
 & 1200500130 & 3 & $6.14\pm0.05$ & $177\pm10$ & $17.8\pm0.6$ & $1.00^{+p}_{-0.04}$ & 0.82\\
 & 1200500131 & F & $5.77\pm0.03$ & $185\pm 4$ & $7.5^{+1.0}_{-0.6}$ & $1.00^{+p}_{-0.3}$ & 0.35\\
 & 1200500134 & 4 & $5.31\pm0.04$ & $176\pm 9$ & $2\pm1$ & $0.30\pm6$ & 0.99\\
 & 1200500135 & F & $4.71\pm0.02$ & $171\pm 3$ & $0.01^{+0.99}_{-p}$ & $0.48^{+0.10}_{-0.01}$ & 0.40\\
 & 1200500137 & F & $5.46\pm0.04$ & $168\pm 4$ & $6.7^{+0.6}_{-2.4}$ & $1.0^{+p}_{-0.4}$ & 0.45\\
 & 1200500140 & F & $5.51\pm0.09$ & $172\pm 8$ & $12\pm3$ & $0.8^{+p}_{-0.1}$ & 0.46\\
 & 1200500142 & F & $5.68\pm0.04$ & $167\pm 4$ & $10.5\pm0.7$ & $1.0^{+p}_{-0.1}$ & 0.58\\
\hline
MAXI~J1803$-$298 & 4202130109 & F & $6.0\pm0.2$ & $198\pm 7$ & $0.004^{+0.269}_{-0.004}$ & $0.21^{+0.04}_{-0.01}$ & 0.66\\
 & 4675020103 & F & $6.88\pm0.08$ & $210\pm 5$ & $0.01^{+0.1}_{-p}$ & $0.23\pm0.03$ & 0.64\\
 & 4202130110 & F & $6.52\pm0.05$ & $195\pm 4$ & $0.01^{+0.02}_{-p}$ & $0.201^{+0.020}_{-0.002}$ & 0.83\\
\hline
MAXI~J1820+070 & 1200120138 & F & $0.178\pm0.004$ & $94\pm 2$ & $142^{+28}_{-41}$ & $0.0^{+0.2}_{-p}$ & 0.78\\
 & 1200120140 & F & $0.250\pm0.005$ & $112\pm 2$ & $61^{+18}_{-2}$ & $0.16^{+0.18}_{-0.09}$ & 0.91\\
 & 1200120141 & F & $0.282\pm0.005$ & $121\pm 3$ & $50^{+60}_{-30}$ & $0.06^{+0.46}_{-p}$ & 0.58\\
 & 1200120142 & F & $0.279\pm0.004$ & $118\pm 1$ & $0.01^{+8}_{-p}$ & $0.26^{+0.30}_{-0.07}$ & 0.63\\
 & 1200120143 & F & $0.299\pm0.004$ & $121\pm 1$ & $7^{+3}_{-1}$ & $0.04^{+0.24}_{-p}$ & 0.68\\
 & 1200120144 & F & $0.311\pm0.003$ & $123\pm 1$ & $16^{+1}_{-2}$ & $0.00^{+0.08}_{-p}$ & 0.66\\
 & 1200120145 & F & $0.338\pm0.003$ & $129\pm 0$ & $0.01^{+53}_{-p}$ & $0.30^{+0.23}_{-0.05}$ & 0.65\\
 & 1200120146 & F & $0.350\pm0.004$ & $129\pm 1$ & $37^{+8}_{-6}$ & $0.09\pm0.06$ & 0.84\\
 & 1200120147 & F & $0.385\pm0.004$ & $131\pm 1$ & $127^{+7}_{-1}$ & $1.00^{+p}_{-0.01}$ & 1.58\\
 & 1200120149 & F & $0.441\pm0.005$ & $133\pm 1$ & $200.0^{+p}_{-0.5}$ & $0.135^{+0.001}_{-0.003}$ & 0.66\\
 & 1200120150 & F & $0.466\pm0.006$ & $138\pm 1$ & $30^{+60}_{-p}$ & $0.0^{+0.4}_{-p}$ & 0.57\\
 & 1200120151 & F & $0.435\pm0.009$ & $134\pm 2$ & $83^{+7}_{-41}$ & $1.00^{+p}_{-p}$ & 0.60\\
 & 1200120152 & F & $0.462\pm0.005$ & $132\pm 1$ & $65\pm6$ & $0.00^{+0.06}_{-p}$ & 1.37\\
 & 1200120153 & F & $0.48\pm0.01$ & $133\pm 4$ & $85^{+5}_{-3}$ & $1.0^{+p}_{-0.4}$ & 1.33\\
 & 1200120156 & F & $0.494\pm0.006$ & $137\pm 1$ & $83\pm4$ & $1.00^{+p}_{-0.04}$ & 0.47\\
 & 1200120157 & F & $0.411\pm0.005$ & $132\pm 1$ & $200.0^{+p}_{-0.5}$ & $0.135\pm0.002$ & 1.62\\
 & 1200120161 & F & $0.335\pm0.004$ & $127\pm 1$ & $38^{+28}_{-36}$ & $0.7\pm0.2$ & 0.64\\
 & 1200120162 & F & $0.329\pm0.007$ & $129\pm 1$ & $30^{+70}_{-p}$ & $0.7\pm0.3$ & 0.68\\
 & 1200120163 & F & $0.423\pm0.007$ & $135\pm 1$ & $0.01^{+3}_{-p}$ & $0.7\pm0.1$ & 1.68\\
 & 1200120165 & F & $0.378\pm0.004$ & $134\pm 1$ & $0.01^{+4}_{-p}$ & $1.0^{+p}_{-0.2}$ & 0.66\\
 & 1200120171 & F & $0.319\pm0.009$ & $129\pm 1$ & $40^{+30}_{-20}$ & $0.1^{+0.2}_{-0.1}$ & 0.59\\
 & 1200120172 & F & $0.337\pm0.007$ & $127\pm 2$ & $0.01^{+23}_{-p}$ & $0.47^{+0.52}_{-0.09}$ & 0.63\\
 & 1200120193 & F & $0.702\pm0.009$ & $142\pm 2$ & $34^{+7}_{-5}$ & $0.00^{+0.05}_{-p}$ & 0.53\\
 & 1200120194 & F & $0.782\pm0.005$ & $148\pm 1$ & $31^{+3}_{-1}$ & $0.28^{+0.08}_{-0.06}$ & 0.74\\
 & 1200120195 & F & $1.57\pm0.01$ & $158\pm 2$ & $9.4^{+0.6}_{-0.4}$ & $0.00^{+0.02}_{-p}$ & 0.56\\
 & 1200120196$^{*}$ & 0 & $2.09\pm0.02$ & $164\pm 3$ & $8^{+0.7}_{-0.5}$ & $0.29\pm0.01$ & 1.05\\
 & 1200120196 & 1 & $2.91\pm0.02$ & $169\pm 7$ & $0.01^{+0.08}_{-p}$ & $0.27\pm0.03$ & 1.25\\
 & 1200120196 & 2 & $3.23\pm0.02$ & $181\pm 6$ & $0.01^{+0.13}_{-p}$ & $0.26\pm0.02$ & 1.18\\
 & 1200120196 & 3 & $3.80\pm0.01$ & $179\pm 7$ & $11^{+9}_{-6}$ & $0.5^{+0.5}_{-0.2}$ & 0.73\\
 & 1200120196 & 4 & $4.16\pm0.03$ & $171\pm 5$ & $0.01^{+0.09}_{-p}$ & $0.23\pm0.02$ & 1.27\\
 & 1200120197 & 2 & $4.99\pm0.05$ & $191\pm 8$ & $0.01^{+0.02}_{-p}$ & $0.20\pm0.01$ & 0.83\\
 & 1200120197 & 11 & $3.93\pm0.02$ & $218\pm 8$ & $3.2\pm0.6$ & $0.17\pm0.01$ & 1.51\\
 & 1200120197 & 12 & $3.25\pm0.07$ & $165\pm 15$ & $1.1^{+3.8}_{-0.7}$ & $0.18\pm0.01$ & 1.21\\
\hline
Swift~J1727.8$-$1613 & 6203980102 & 0 & $0.276\pm0.007$ & $87\pm 15$ & $0.01^{+35}_{-p}$ & $0.72\pm0.06$ & 1.08\\
 & 6203980102 & 1 & $0.324\pm0.006$ & $80\pm 15$ & $49^{+1}_{-15}$ & $0.02^{+0.05}_{-p}$ & 1.10\\
 & 6203980102 & 2 & $0.345^{+0.004}_{-0.007}$ & $91\pm 6$ & $200^{+p}_{-10}$ & $1.00^{+p}_{-0.04}$ & 1.07\\
 & 6203980102$^{*}$ & 3 & $0.371\pm0.004$ & $105\pm 5$ & $0.01^{+21}_{-p}$ & $0.54\pm0.01$ & 1.42\\
 & 6203980103 & 2 & $0.455\pm0.008$ & $99\pm 14$ & $0.01^{+15}_{-p}$ & $0.8\pm0.1$ & 1.20\\
 & 6203980103 & 3 & $0.597\pm0.005$ & $114\pm 5$ & $0.01^{+32}_{-p}$ & $0.56\pm0.05$ & 1.32\\
 & 6203980104 & F & $0.795\pm0.005$ & $120\pm 2$ & $29\pm2$ & $0.28^{+0.10}_{-0.06}$ & 1.46\\
 & 6203980105 & F & $0.891\pm0.004$ & $125\pm 1$ & $20\pm2$ & $0.23\pm0.02$ & 1.61\\
 & 6203980109$^{*}$ & F & $1.266\pm0.002$ & $137\pm 1$ & $13.21\pm0.02$ & $0.331\pm0.003$ & 1.62\\
 & 6203980110 & F & $1.315\pm0.003$ & $135\pm 1$ & $13.05\pm0.03$ & $0.345\pm0.005$ & 1.68\\
 & 6203980111$^{*}$ & F & $1.177\pm0.003$ & $138\pm 1$ & $15.5\pm0.3$ & $0.32\pm0.02$ & 1.32\\
 & 6203980112 & 0 & $1.030\pm0.006$ & $140\pm 3$ & $44^{+0.4}_{-0.6}$ & $1.00^{+p}_{-0.02}$ & 1.62\\
 & 6203980112 & 1 & $1.10\pm0.01$ & $132\pm 6$ & $8\pm2$ & $0.7\pm0.1$ & 1.29\\
 & 6203980112 & 2 & $1.20\pm0.01$ & $132\pm 9$ & $35^{+2}_{-11}$ & $1.0^{+p}_{-0.4}$ & 1.31\\
 & 6203980112 & 3 & $1.32\pm0.03$ & $142\pm 5$ & $17^{+4}_{-6}$ & $0.4\pm0.2$ & 1.42\\
 & 6203980112 & 4 & $1.49\pm0.01$ & $138\pm 5$ & $6\pm2$ & $0.37\pm0.04$ & 1.59\\
 & 6203980112 & 6 & $1.43\pm0.01$ & $147\pm 4$ & $13\pm5$ & $0.39^{+0.17}_{-0.09}$ & 1.58\\
 & 6203980112 & 8 & $1.396\pm0.009$ & $149\pm 6$ & $4\pm1$ & $0.37\pm0.05$ & 1.56\\
 & 6203980112 & 10 & $1.492^{+0.006}_{-0.009}$ & $144\pm 3$ & $8^{+1}_{-2}$ & $0.32\pm0.02$ & 1.69\\
 & 6203980113 & F & $1.400\pm0.004$ & $145\pm 1$ & $13.77^{+0.06}_{-0.08}$ & $0.41\pm0.03$ & 1.66\\
 & 6203980114$^{*}$ & F & $1.377\pm0.005$ & $146\pm 1$ & $13.61\pm0.03$ & $0.399\pm0.006$ & 1.47\\
 & 6703010104 & F & $1.62\pm0.01$ & $136\pm 5$ & $0.01^{+3}_{-p}$ & $0.54\pm0.01$ & 0.88\\
 & 6703010107$^{*}$ & F & $2.56\pm0.01$ & $177\pm 3$ & $2.02\pm0.08$ & $0.471^{+0.002}_{-0.004}$ & 1.34\\
 & 6203980130 & F & $6.5\pm0.1$ & $171\pm 7$ & $2.6\pm0.5$ & $0.4\pm0.1$ & 1.09\\
 & 6703010113$^{*}$ & F & $7.4\pm0.1$ & $143\pm 3$ & $6.6\pm0.3$ & $0.00^{+0.02}_{-p}$ & 0.71\\
 & 6557020401$^{*}$ & F & $7.3\pm0.1$ & $162\pm 2$ & $2.9\pm0.2$ & $0.71\pm0.07$ & 0.48\\
\hline
\caption*{* Observations with significant \nicer\ calibration systematics. \\
** Observations with significant wind absorption lines.}
\end{longtable}

\twocolumn

\section{Samples of joint fitting}\label{sec:appendix samples}

\begin{figure*}
    \includegraphics[width=0.88\textwidth,trim=10 25 100 80]{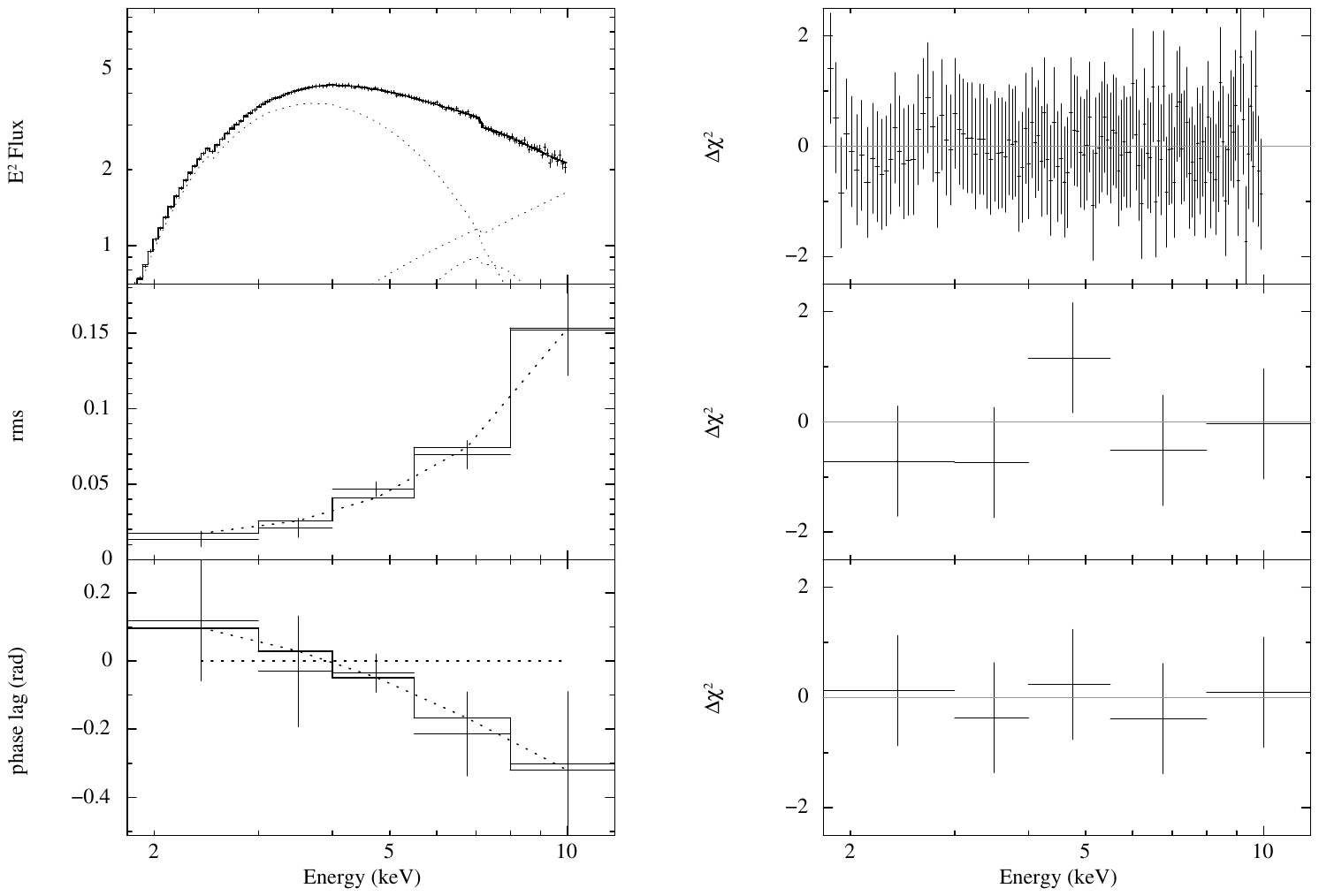}
    \caption{Example fit for EXO~1846$-$031 (Observation ID 2200760110). Left: from top to bottom, the time-averaged energy spectrum of the source, rms and phase lag spectra of the QPO, along with the corresponding best-fit models. Right: from top to bottom, the residuals ($\Delta \chi^2$) for each of the corresponding fits shown on the left.}\label{fig:fit}
\end{figure*}
\begin{figure*}
    \includegraphics[width=0.88\textwidth,trim=10 25 100 80]{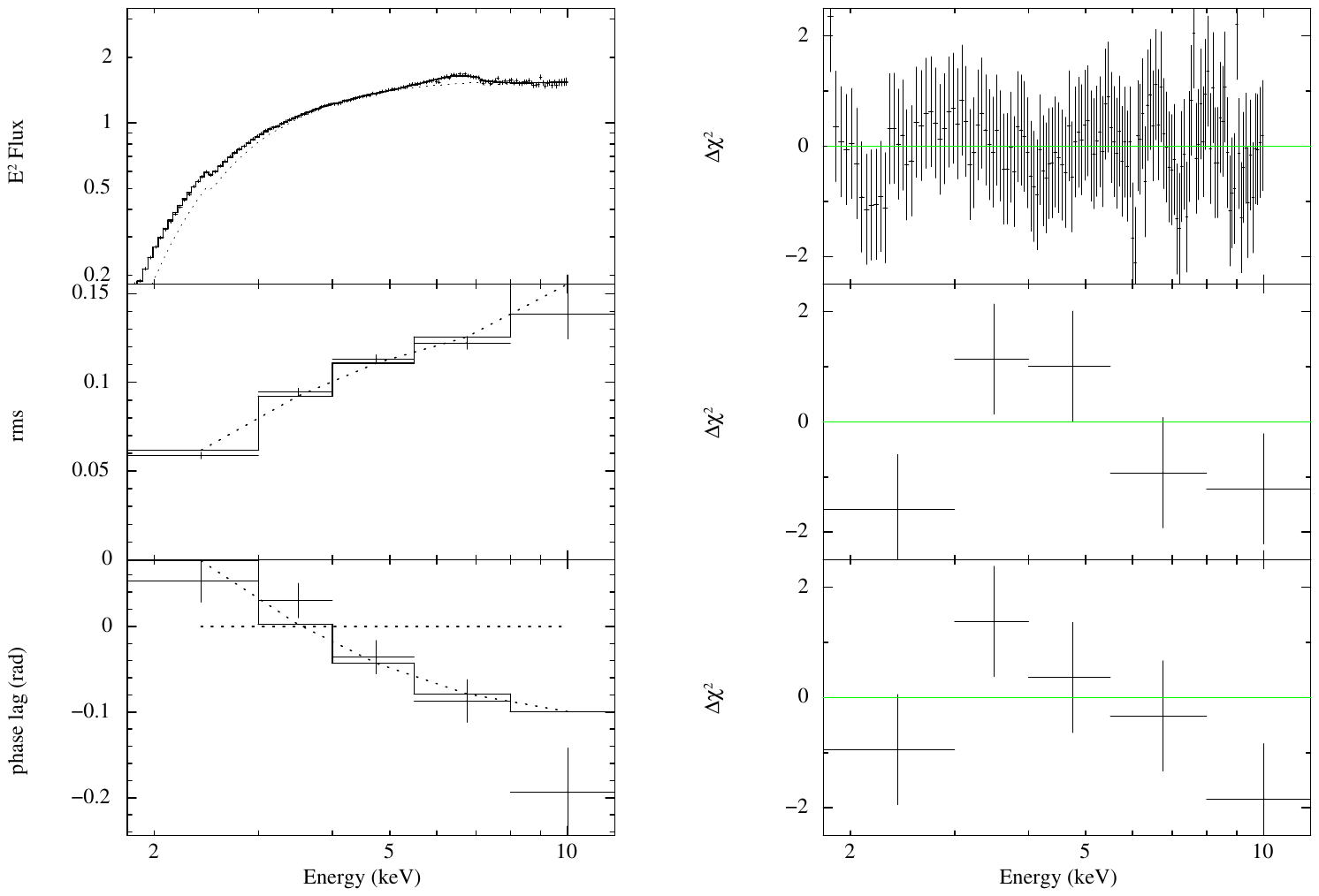}
    \caption{Example fit for GRS~1915+105 (Observation ID 1103010157). Panels are the same as in Fig.~\ref{fig:fit}.}
\end{figure*}
\begin{figure*}
    \includegraphics[width=0.88\textwidth,trim=10 25 100 80]{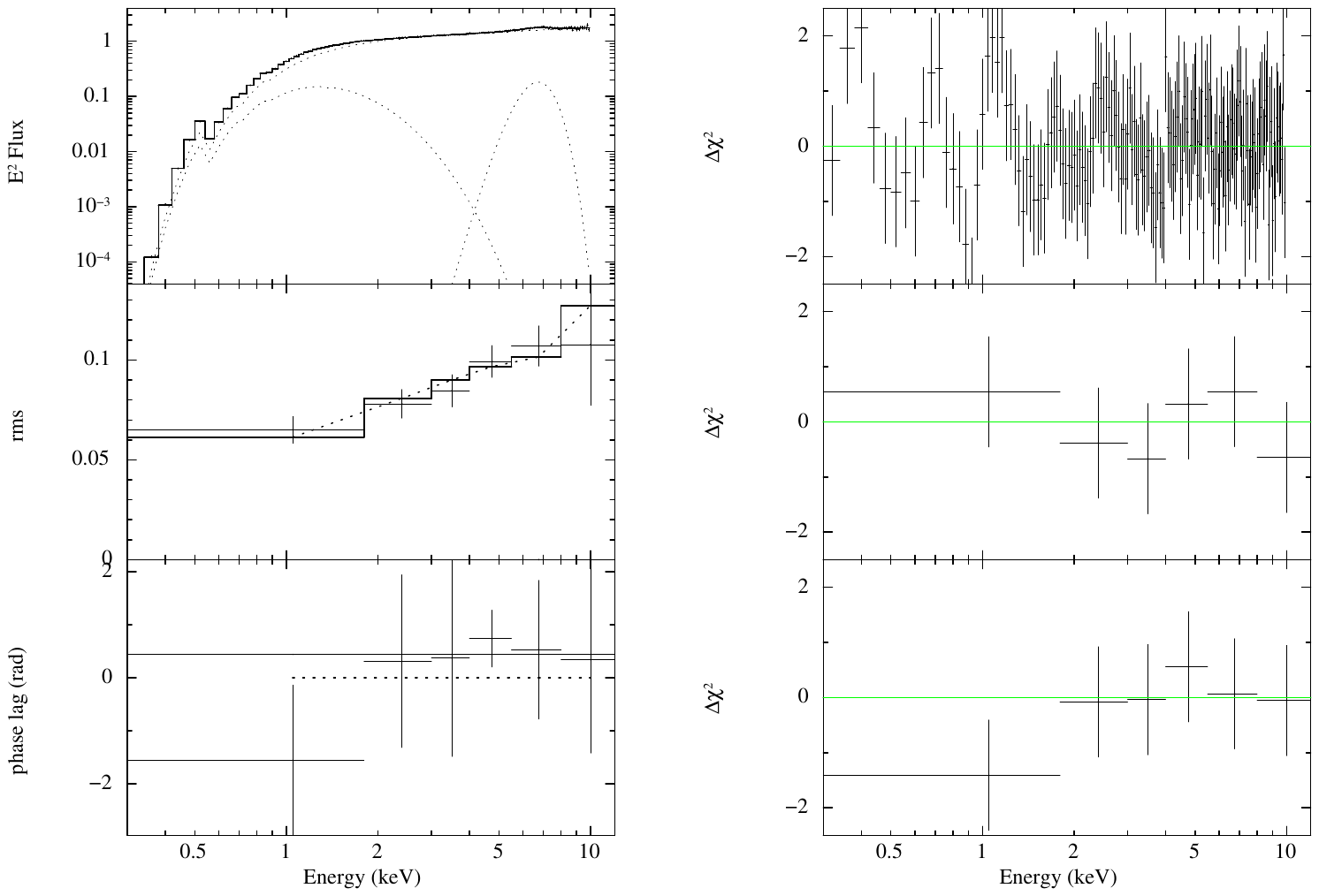}
    \caption{Example fit for GX~339$-$4 (Observation ID 4133010101). Panels are the same as in Fig.~\ref{fig:fit}.}
\end{figure*}
\begin{figure*}
    \includegraphics[width=0.88\textwidth,trim=10 25 100 80]{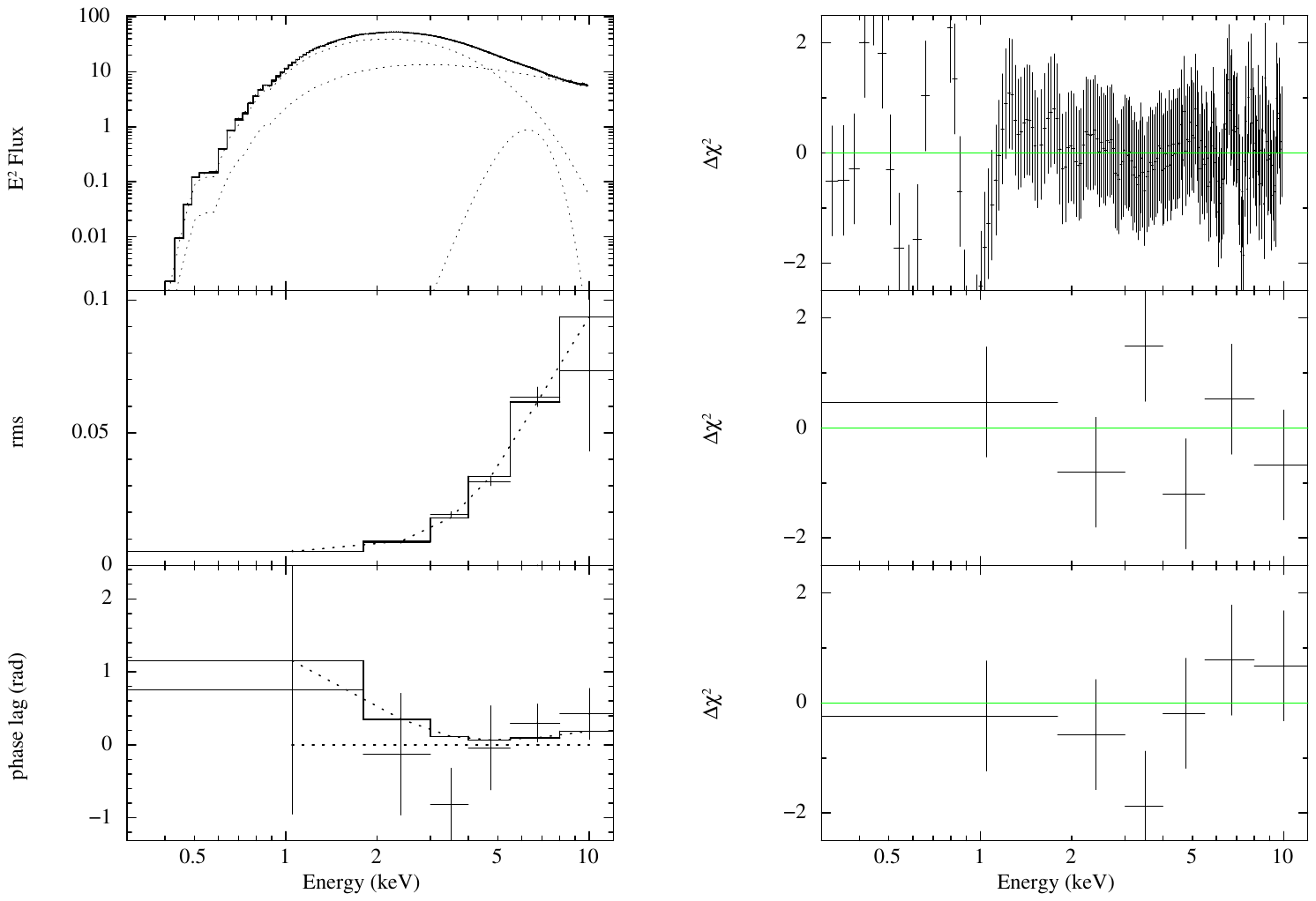}
    \caption{Example fit for MAXI~J1348$-$630 (Observation ID 1200530113). Panels are the same as in Fig.~\ref{fig:fit}.}
\end{figure*}
\begin{figure*}
    \includegraphics[width=0.88\textwidth,trim=10 25 100 80]{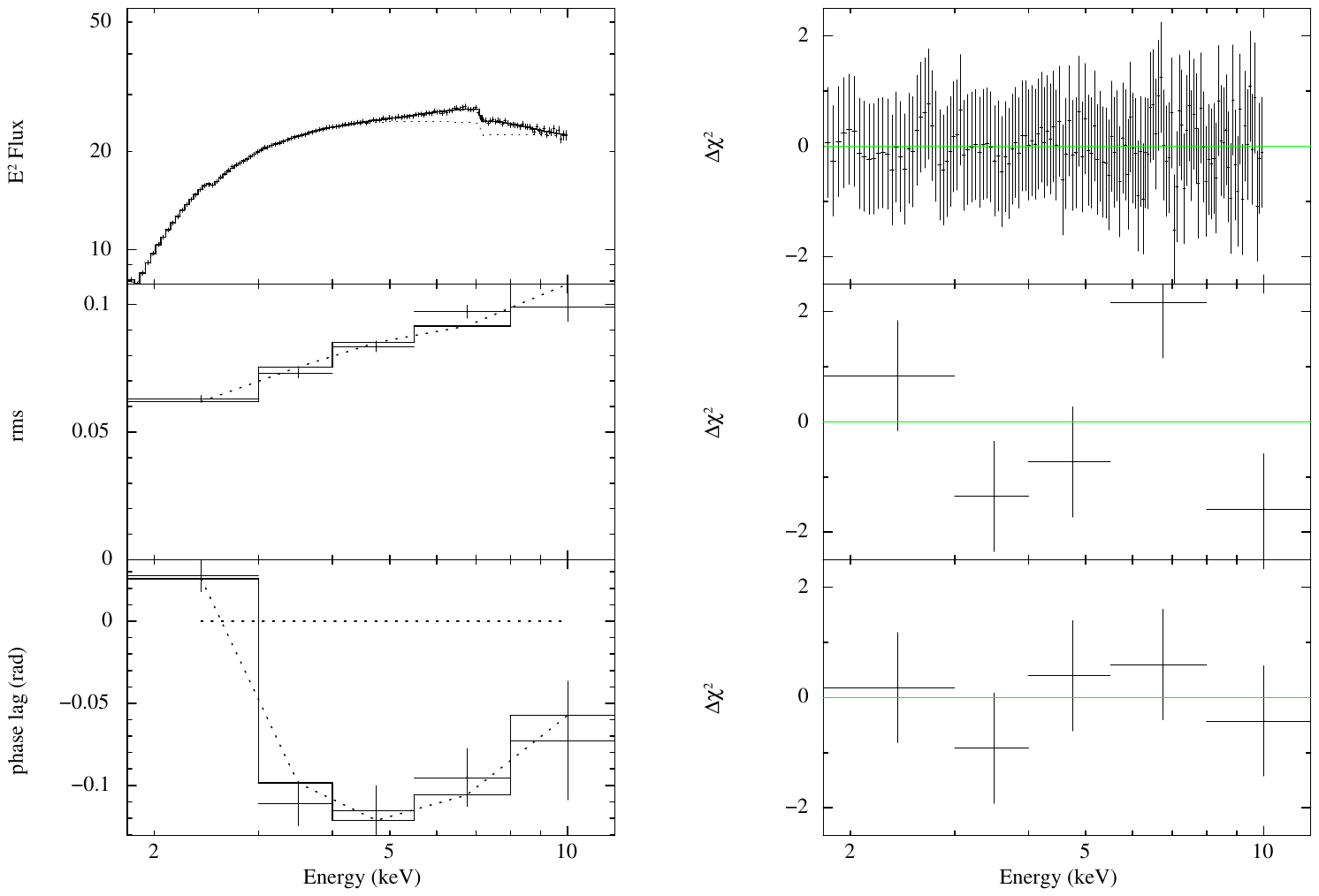}
    \caption{Example fit for MAXI~J1535$-$571 (Observation ID 1050360107). Panels are the same as in Fig.~\ref{fig:fit}.}
\end{figure*}
\begin{figure*}
    \includegraphics[width=0.88\textwidth,trim=10 25 100 80]{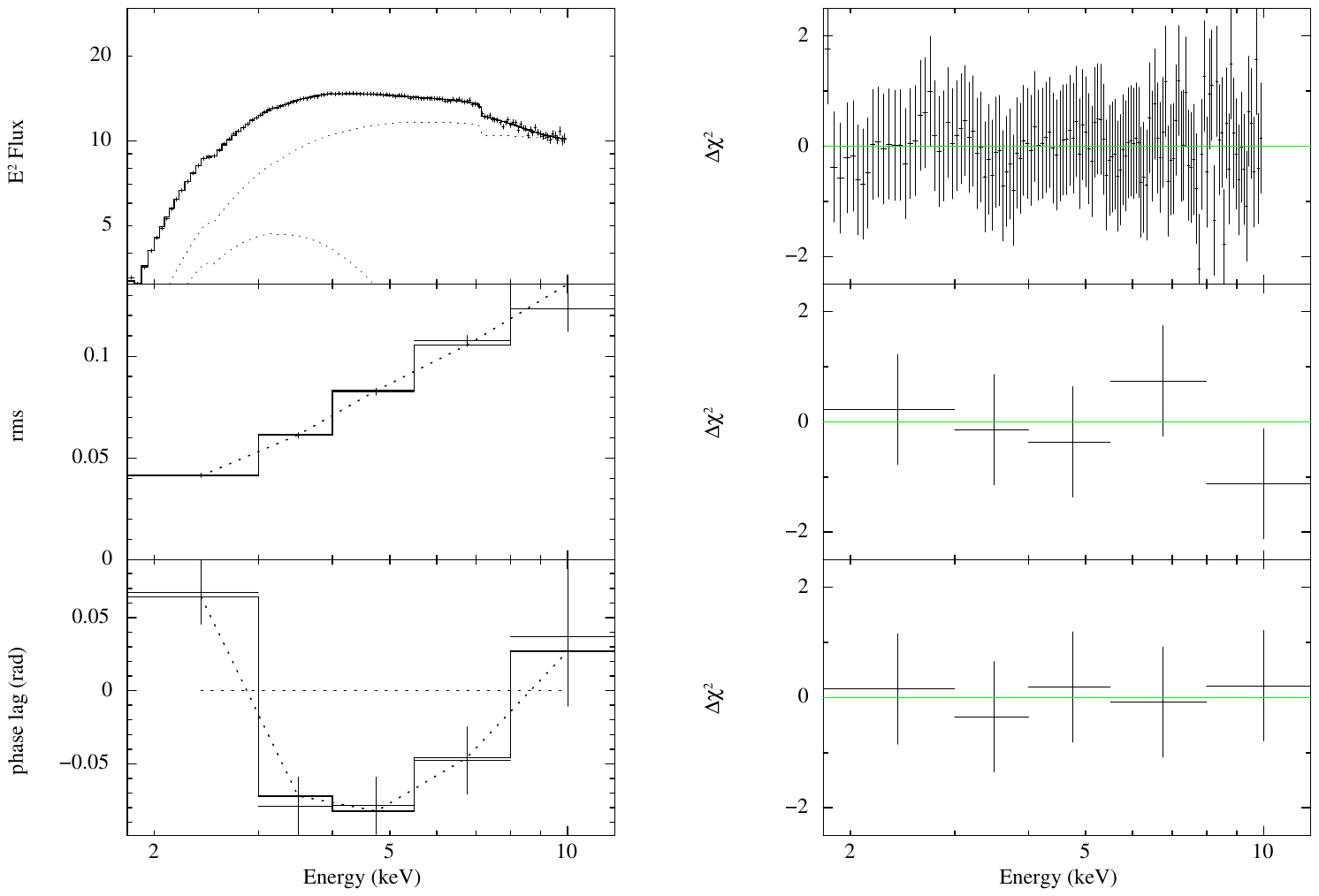}
    \caption{Example fit for MAXI~J1631$-$479 (Observation ID 1200500111). Panels are the same as in Fig.~\ref{fig:fit}.}
\end{figure*}
\begin{figure*}
    \includegraphics[width=0.88\textwidth,trim=10 25 100 80]{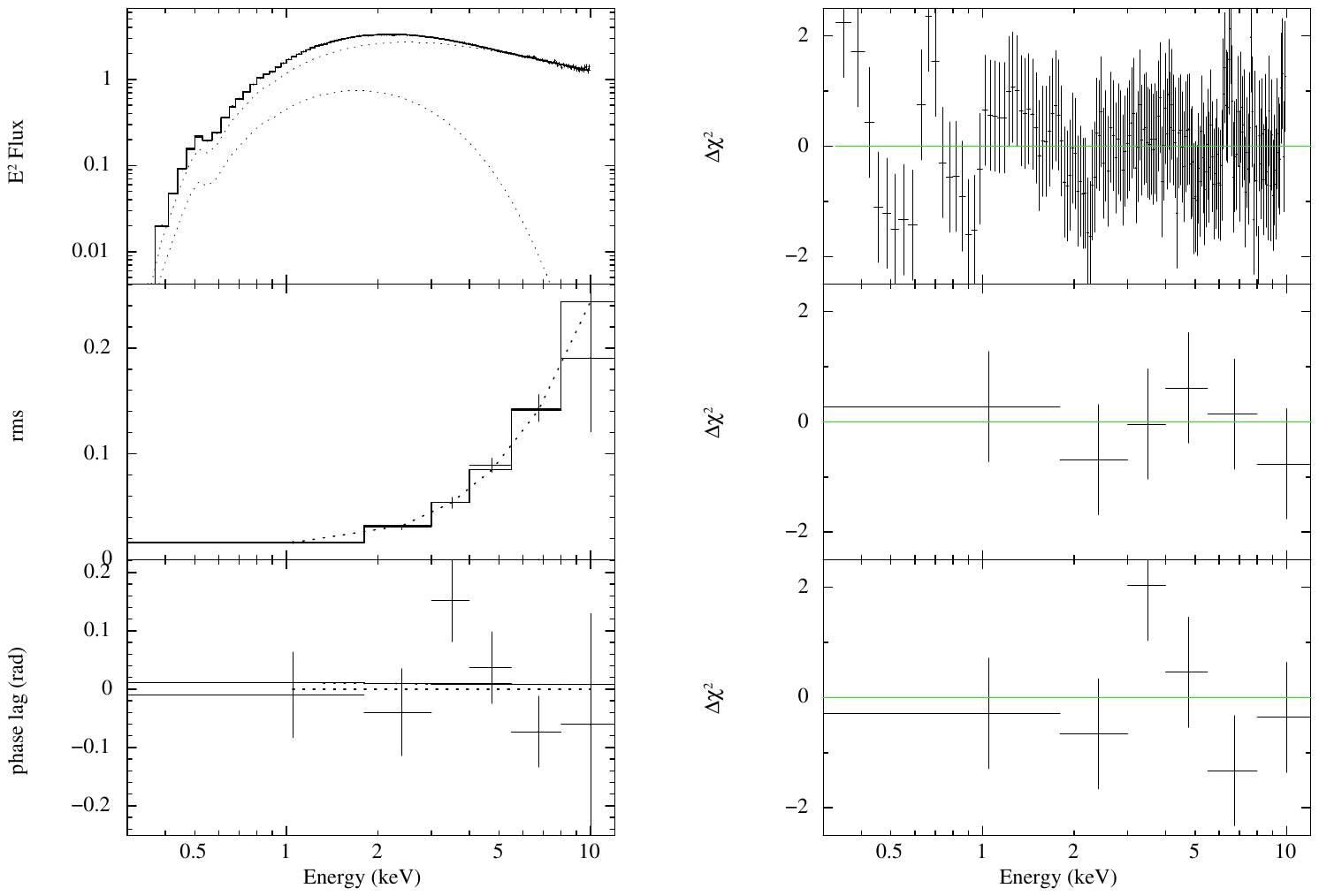}
    \caption{Example fit for MAXI~J1803$-$298 (Observation ID 4202130109). Panels are the same as in Fig.~\ref{fig:fit}.}
\end{figure*}
\begin{figure*}
    \includegraphics[width=0.88\textwidth,trim=10 25 100 80]{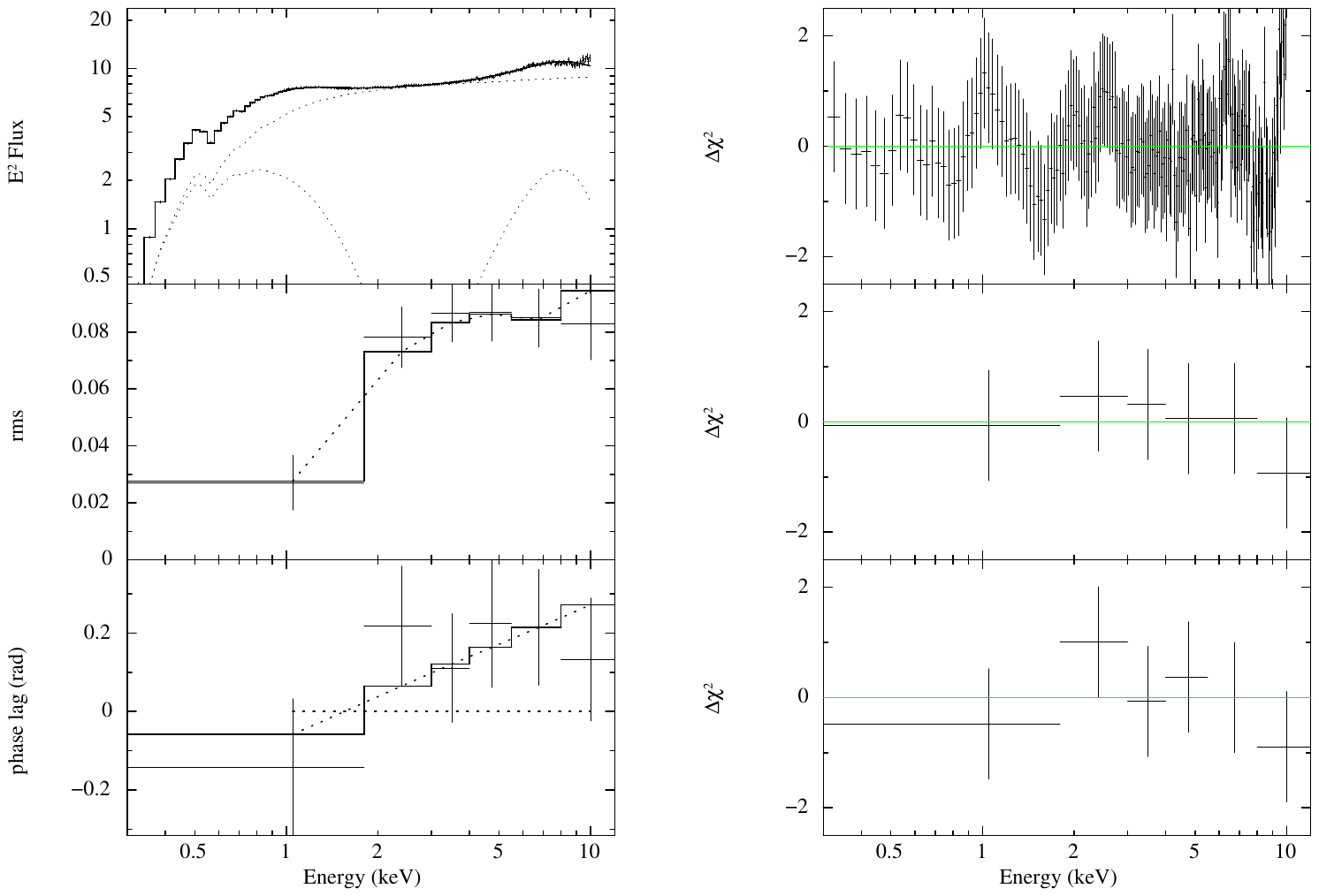}
    \caption{Example fit for MAXI~J1820+070 (Observation ID 1200120141). Panels are the same as in Fig.~\ref{fig:fit}.}
\end{figure*}
\begin{figure*}
    \includegraphics[width=0.88\textwidth,trim=10 25 100 80]{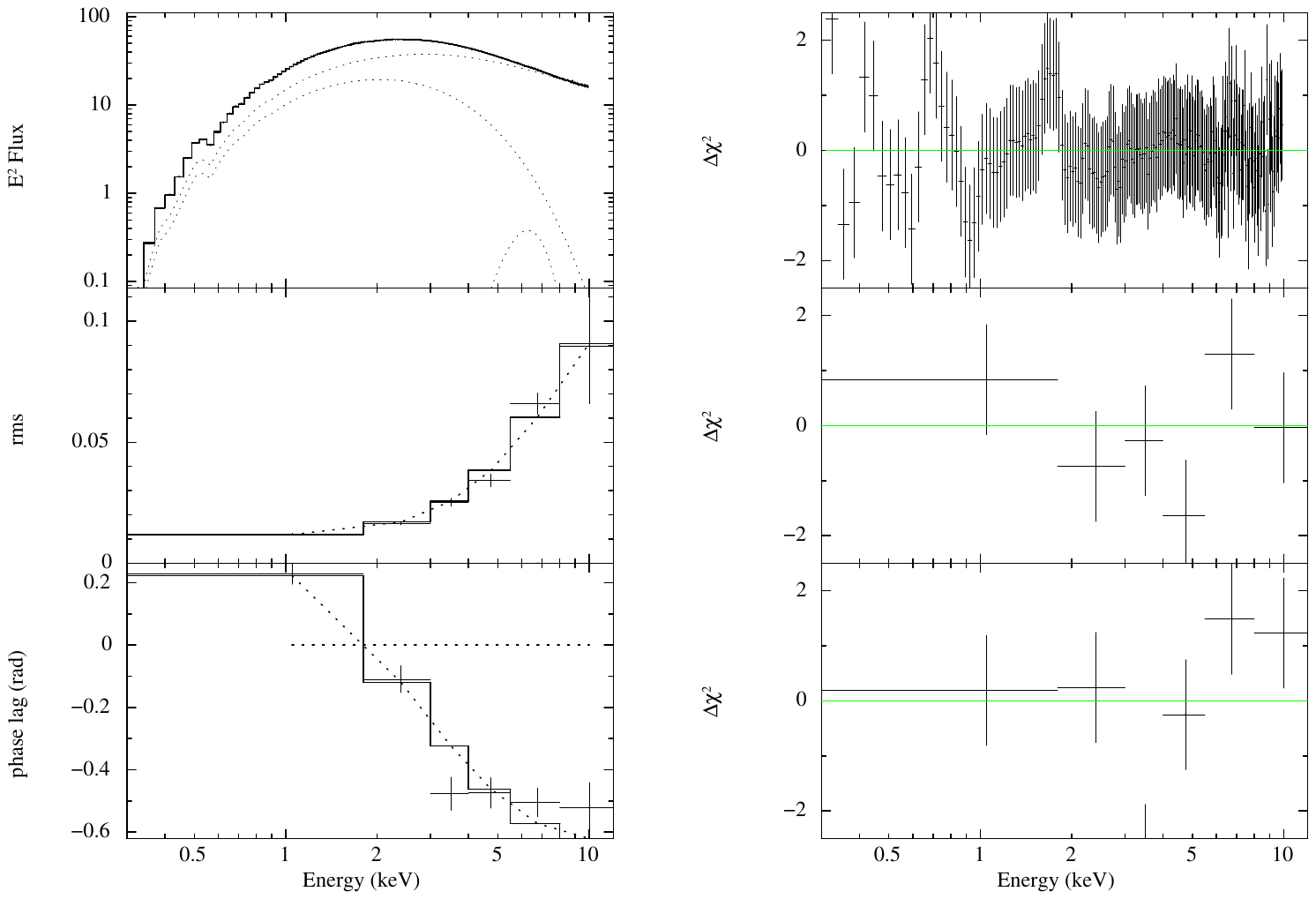}
    \caption{Example fit for Swift~J1727.8$-$1613 (Observation ID 6557020401). Panels are the same as in Fig.~\ref{fig:fit}.}
\end{figure*}


\bsp	
\label{lastpage}
\end{document}